\documentclass[aps,prd,nofootinbib,reprint,superscriptaddress]{revtex4-1}

\usepackage{hyperref}
\usepackage{textcomp}
\usepackage{blindtext}
\usepackage{amsfonts}
\usepackage{tensor}
\usepackage{graphicx}
\usepackage{pict2e}
\usepackage{balance}
\usepackage{ulem}
\usepackage[T1]{fontenc}
\usepackage[utf8]{inputenc}

\usepackage{amssymb, amsmath,environ}

\usepackage{color}


\newcommand{\bse}{\begin{subequations}}
	\newcommand{\ese}{\end{subequations}}
\newcommand{\be}{\begin{equation}}
\newcommand{\ee}{\end{equation}}

\NewEnviron{eqsplit}{%
	\begin{equation}\begin{split}
	\BODY
	\end{split}\end{equation}
}

\makeatletter
\newcommand*\bigcdot{\mathpalette\bigcdot@{.5}}
\newcommand*\bigcdot@[2]{\mathbin{\vcenter{\hbox{\scalebox{#2}{$\m@th#1\bullet$}}}}}
\makeatother

\newcommand{\bea}{\begin{eqnarray}}
\newcommand{\eea}{\end{eqnarray}}
\newcommand{\ba}{\begin{array}}
	\newcommand{\ea}{\end{array}}

\newcommand{\la}{\langle}
\newcommand{\ra}{\rangle}

\begin{document}

\title{Longitudinal flow decorrelations in light ion collisions}

\begin{abstract}
	This study presents a detailed analysis of longitudinal flow decorrelations in light ion collisions at Relativistic Heavy Ion Collider (RHIC) and Large Hadron Collider(LHC) energies for the first time. We compare different theoretical models of the oxygen structure and find that certain observables such as the flow vector correlator can distinguish between them, particularly highlighting differences between the variational Monte Carlo (VMC) structure and other models, such as Nuclear Lattice Effective Field Theory (NLEFT) and the Projected Generator Coordinate Method (PGCM), which behave similarly.
	We further examine flow decorrelations across different systems, including d+Au and O+O collisions at 200 GeV, as well as Ne+Ne and O+O collisions at 6.37 TeV, indicating a hierarchy in decorrelations that underscores the complexity of these interactions. Our findings emphasize that the role of asymmetry in d+Au collisions and its impact on flow correlations is mitigated using a modified flow correlation defined as a ratio of two flow vector covariances derived from different pairs of pseudorapidity bins. Additionally, our analysis of flow angle decorrelations shows diverse distributions across various systems, finding distinct patterns in d+Au collisions compared to other collision systems.
\end{abstract}

\author{Hadi Mehrabpour}
\email{mehrabpour@pku.edu.cn}
\affiliation{School of Physics, Peking University, Beijing 100871, China}
\affiliation{Center for High Energy Physics, Peking University, Beijing 100871, China}

\author{Abhisek Saha}
\email{saha@pku.edu.cn}
\affiliation{School of Physics, Peking University, Beijing 100871, China}
\affiliation{Center for High Energy Physics, Peking University, Beijing 100871, China}

\maketitle

\section{Introduction}\label{Introduction}
The study of nuclear structure has long been a cornerstone of nuclear physics, providing insights into the fundamental forces and interactions that govern matter at the subatomic level \cite{frank}. Investigating nuclear structure in relation to the complex nucleon-nucleon interactions presents a significant challenge in low-energy nuclear physics \cite{Delaroche:2009fa,Wang:2024kdo}, driving research into various phenomenological models \cite{Wang:2024kdo,Demyanova:2024ahe,Hamada:2023sjd,Morales-Gallegos:2022dzq,Kekejian:2022ipn} and experimental methodologies \cite{Yang:2022wbl,Magdy:2024thf,Cline:1986ik}. This complexity has been explored through advanced models, including NLEFT simulations \cite{Meissner:2014lgi,Elhatisari:2017eno}, PGCM \cite{Frosini:2021fjf,Frosini:2021sxj,Frosini:2021ddm}, and VMC \cite{Lonardoni:2018nob}, in nuclei such as $^{16}$O and $^{20}$Ne.
The effort of low-energy experiments is complemented by exploring nuclear structure in ultra-relativistic ion collisions \cite{Jia:2022ozr}. These investigations rely on nucleon distribution data derived from various theoretical models.

In contemporary nuclear physics, relativistic nuclear collisions at the Relativistic Heavy Ion Collider (RHIC) and the Large Hadron Collider (LHC) stand out as particularly effective methodologies, allowing for a comprehensive analysis of the structural properties associated with colliding nuclei. The matter distribution within these colliding nuclei is illuminated by the shape and size of the energy density in the overlap region \cite{STAR:2024wgy,Giacalone:2023hwk,Ollitrault:2023wjk}. The initial energy deposition is influenced by both the collision geometry and the quantum fluctuations present in the nuclear wavefunctions \cite{Alver:2010gr}, resulting in event-by-event fluctuations \cite{Stephanov:1999zu,Aguiar:2001ac}. It is well-established that the initial state properties are analyzed through collective flow obtained from the spectra of emitted particles \cite{Huovinen:2006jp,Voloshin:2008dg,Heinz:2013th}. Since flow is sensitive to matter distribution in nuclei \cite{Jia:2021qyu,Jia:2021tzt,Fortier:2024yxs, Bally:2021qys, Zhang:2021kxj,Giacalone:2021udy,Nijs:2021kvn,Jia:2022qgl}, it serves as a suitable observable for studying nuclear structures. Recently, a number of studies have been conducted to investigate light nuclear structures, particularly $^{16}O$ and $^{20}Ne$, which may pave the way for discoveries in high-energy experiments \cite{Jia:2022ozr}. 
In Ref.\cite{Summerfield:2021oex}, the NLEFT model 
has been utilized to generate the oxygen nucleonic configurations. The first measurements of azimuthal anisotropies in O+O collisions for the collected data at RHIC have been done in Ref.\cite{Huang:2023viw}.  The results showed that the PHOBOS Glauber model \cite{Alver:2008aq} based on the VMC can predict the experimental data to a good extent. Moreover, \textit{ab-initio} nucleon-nucleon correlations (VMC, NLEFT, PGCM) and their impact on O+O collisions have been studied at RHIC energies \cite{Zhang:2024vkh}.  When comparing the ratio of elliptic flow obtained from VMC with respect to the nucleon configurations derived by starting from a three parameter Fermi (3pF) density distribution, they found a large deviation in central collisions compared to NLEFT and PGCM. Since the advanced  models are able to address the emergence of the clustering correlations in light nuclei, Refs.\cite{Mehrabpour:2025ogw,YuanyuanWang:2024sgp,Prasad:2024ahm} have explored the effects of $\alpha$-clustering in the relativistic O+O collisions.  A comparison of the $^{16}$O and $^{20}$Ne structures has been performed for both symmetric and asymmetric collision systems \cite{Mehrabpour:2025ogw,Giacalone:2024luz,Giacalone:2024ixe}. The bowling pin shape of Ne in comparison with oxygen structures has been studied.

 The studies have so far been done on the high energy light nuclear collisions were constraint to the transverse dynamics of the systems.
The rapid expansion of the fireball due to the presence of strong pressure gradients, information about initial fluctuations is also encoded in the longitudinal direction \cite{Bozek:2010vz,Broniowski:2011jm,Bozek:2015swa,Bozek:2015bna,Bzdak:2015dja,Bozek:2017thv,Bozek:2022cjj}. Investigations utilizing models to analyze two-particle correlations as a function of pseudorapidity have uncovered significant event-by-event fluctuations in both the magnitude and phase of the flow. This is particularly evident between two distinct pseudorapidities, leading to notable discrepancies such that $v_n(\eta_a)$ differs from $v_n(\eta_b)$, indicative of forward-backward asymmetry, and $\psi_n(\eta_a)$ varies from $\psi_n(\eta_b)$, reflecting a twisting of the event plane \cite{Bozek:2010vz,Xiao:2012uw,Jia:2014ysa}.  Recently, flow decorrelations have been analyzed in collisions involving heavy nuclei of similar masses but different structures \cite{Zhang:2024bcb}, spherical or deformed shapes, using a dynamical transport model. By reducing the impact of non-flow effects, it has been found that while non-spherical shapes, such as prolate or oblate shapes characterized by quadrupole deformation parameter $\beta_2$, considerably increase the overall flow magnitude $v_2$, they do not change its longitudinal profile.

To present a comprehensive study of small systems, investigation of light nuclear structure in the longitudinal direction is required. The aim of this paper is to investigate  the effects of light nuclear structures on the longitudinal flow decorrelations for the recently proposed small collision systems O+O and Ne+Ne, as well as d+Au. In Section \ref{sec2}, we provide a brief overview of the materials required for this analysis.  The variations in Hamiltonians and approximations lead to significant differences in the structural predictions of light nuclei, as resulted by the calculations from stated theories. We investigate how these differences lead to divergent predictions for the observables of oxygen in the longitudinal direction in Sec.\ref{sec3a}. Within this section, we focus on the differences arising from spherical oxygen configurations by isolating the deformation components. In Section \ref{sec3b}, we conduct a comparative analysis of O+O collisions and d+Au collisions to explore nuclear structures at different scales within the context of RHIC energy. We also present predictions regarding longitudinal decorrelation for O+O and Ne+Ne collisions in anticipation of the upcoming LHC Run. The findings are summarized in the concluding section.     

\begin{figure}[t!]
	\begin{center}
		\begin{tabular}{c}
			\hspace*{-1.cm}\includegraphics[scale=0.65]{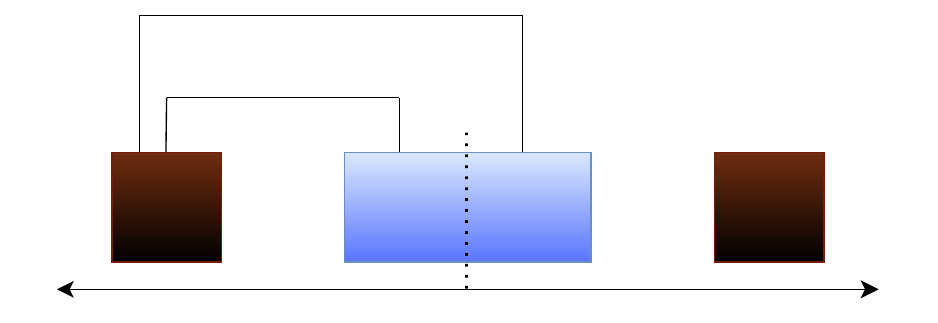}	
		\end{tabular}
		\begin{picture}(0,0)
		\put(-100,35){{\fontsize{15}{15}\selectfont \textcolor{white}{B}}}
		\put(90,35){{\fontsize{15}{15}\selectfont \textcolor{white}{F}}}
		\put(-35,35){{\fontsize{15}{15}\selectfont \textcolor{red}{$-\eta$}}}
		\put(10,35){{\fontsize{15}{15}\selectfont \textcolor{red}{$\eta$}}}
		\put(-95,80){{\fontsize{9}{9}\selectfont \textcolor{black}{$\la V(-\eta_{\text{ref}})V^*(-\eta)\ra$}}}
		\put(-70,105){{\fontsize{9}{9}\selectfont \textcolor{black}{$\la V(-\eta_{\text{ref}})V^*(\eta)\ra$}}}
		\end{picture}		
		\caption{This is a schematic of the longitudinal or beam direction. It shows the use of pseudorapidity bins in the definition of longitudinal decorrelations.} 
		\label{fig1}
	\end{center}
\end{figure}

\section{Materials}\label{sec2}
In the study of high-energy nuclear collisions,  the anisotropic flow is pivotal for understanding the azimuthal distribution of emitted particles \cite{Huovinen:2006jp,Voloshin:2008dg,Heinz:2013th}. Flow can be quantitatively described using harmonic coefficients, specifically through the notation $V_n = v_n e^{i2\psi_n}$ \cite{Voloshin:1994mz,Luzum:2013yya,ATLAS:2012at}, where $v_n$ represents the magnitude of harmonic flows and  $\psi_n$  denotes the event-plane angle associated with them. Here, the investigation focuses on elliptic flow $V_2$, which reflects the hydrodynamic response to the elliptically-shaped overlap region and its associated quadrupole deformations. 
Nevertheless, it is crucial to recognize that the flow vector $V_2$ cannot be derived from isolated events \cite{Mehrabpour:2018kjs}. Rather, we drive rotationally invariant combinations of flow vectors by analyzing the moments of the associated $q_2$ vectors within a defined region of phase space $\eta_a$: $q_2(\eta_a)=\frac{1}{N}\sum_{k \in \eta_a} e^{i2\phi_k}$ \cite{Bilandzic:2010jr,ATLAS:2012at}, where the sum is performed over all  $N$ hadrons located within the specified phase space region  $\eta_a$ , and $\phi_k$ represents the azimuthal angles of these particles \cite{Bozek:2023dwp}. The event average of the  $q_2$ vector moments acts as an estimator for the corresponding moments of the flow vectors: $V_{2\Delta}(\eta_a,\eta_b)\equiv\langle V_2(\eta_a)V_2^*(\eta_b)\rangle =  \langle  q_2(\eta_a) q_2^*(\eta_b) \rangle$, where the angular brackets denote an average taken over multiple events \cite{Bozek:2015tca}. The moments $V_{2\Delta}(\eta_a,\eta_b)$  can also reflect the phenomenon of factorization breaking in collective flow\footnote{Factorization breaking not only predicts a deviation from unity in the correlation coefficient between two flow vectors measured in separate pseudorapidity region, but it also encompasses this prediction between two bins of transverse momentum.}\cite{Bozek:2010vz}. This indicates that flow moments derived from different phase space regions  $\eta_a$  and  $\eta_b$ is the following \cite{Bozek:2010vz,Gardim:2012im}:
\begin{equation}\label{q1}
\begin{split}
 v_2\{2\}^2(\eta_a,\eta_b)&=V_{2\Delta}(\eta_a,\eta_b)\\& = \frac{1}{N_{events}}\sum_{events} \frac{1}{N_a N_b}\sum_{k \in \eta_a, j\in \eta_b } e^{i 2(\phi_k-\phi_j)}.
\end{split}
\end{equation}
In contrast, the formula for flow moments calculated within a single region is represented as follows: $$ v_2\{2\}^2(\eta_a) = \frac{1}{N_{events}}\sum_{events} \frac{1}{N_a (N_a-1)}\sum_{k\neq j  \in \eta_a } e^{i 2(\phi_k-\phi_j)}.$$ 
To measure the extent of decorrelation between flow vectors in two distinct regions of phase space, we utilize the decorrelation coefficient \cite{CMS:2015xmx,ATLAS:2017rij,CMS:2013bza,Zhou:2014bba}:
\begin{equation}\label{q2}
R(\eta_a,\eta_b)=\frac{ v_2\{2\}^2(\eta_a,\eta_b)}{\sqrt{ v_2\{2\}^2(\eta_a) v_2\{2\}^2(\eta_b)}}.
\end{equation}
In scenarios where multiparticle correlations are primarily influenced by flow, we find that  $R(\eta_a,\eta_b) < 1$. This correlation coefficient serves as a valuable measure of flow decorrelation across different pseudorapidity bins \cite{Bozek:2010vz}.

The decorrelation observed in pseudorapidity is significantly affected by non-flow effects \cite{CMS:2015xmx}. To tackle this issue, a modified factorization breaking coefficient has been introduced, defined as a ratio of two flow vector covariances derived from different pairs of bins \cite{CMS:2015xmx,Huo:2017hjv,Bozek:2015bha,Xiao:2015dma,Jia:2017kdq,Bozek:2018nne}:
\begin{align}
r_{n;V}(\eta)\equiv r_{n;V}(\eta,\eta_{ref}) &=\frac{v_n\{2\}^2(\eta,-\eta_{ref})}{v_n\{2\}^2(-\eta,-\eta_{ref})},
\label{eq:Rv}\\
r_{n;\psi}(\eta)\equiv r_{n;\psi}(\eta,\eta_{ref})&=\frac{\langle\cos\left(n \psi_n(\eta)-n\psi_n(-\eta_{ref})\right) \rangle }{\langle\cos\left(n\psi_n(-\eta)-n\psi_n(-\eta_{ref})\right) \rangle }.
\label{eq:Rc}
\end{align}
In this context, $\eta_{ref}$ acts as a common reference pseudorapidity, as illustrated in Fig.\ref{fig1}. Experimental results indicate that decorrelation described by Eqs.\ref{eq:Rv} and \ref{eq:Rc} can be qualitatively replicated by hydrodynamic and cascade models \cite{CMS:2015xmx,ATLAS:2017rij}. This provides a robust framework for elucidating these indicated phenomena in high-energy nuclear collisions. 

In the following sections, we explore longitudinal decorrelations in small systems at RHIC and LHC energy collisions. To achieve this, we employ a combination of the 3D-Glauber Monte-Carlo model \cite{Shen:2017bsr}, (3+1)D viscous hydrodynamics (MUSIC)\cite{Schenke:2010rr}, and UrQMD \cite{Schenke:2010nt} to generate 10000 fluctuating events for each type of collisions: d+Au and O+O at $200$ GeV, as well as O+O and Ne+Ne at $6.37$ TeV in preparation for the upcoming LHC Run. We consider 2000 oversampled events derived from the same hydrodynamic event, and then they are combined in the hadronic flow analysis.  It is essential to emphasize that we have established the centrality range at 0-5\%, where we expect to observe the maximum deviation in flow correlations for various \textit{ab initio} models \cite{Zhang:2024vkh}. Additionally, we have chosen a transverse momentum range of $0.2 \leq p_T\leq 3$ GeV to improve the robustness of our statistical sample.
\begin{figure}[t!]
	\begin{tabular}{c}
		\hspace*{.7cm}\includegraphics[scale=1.1]{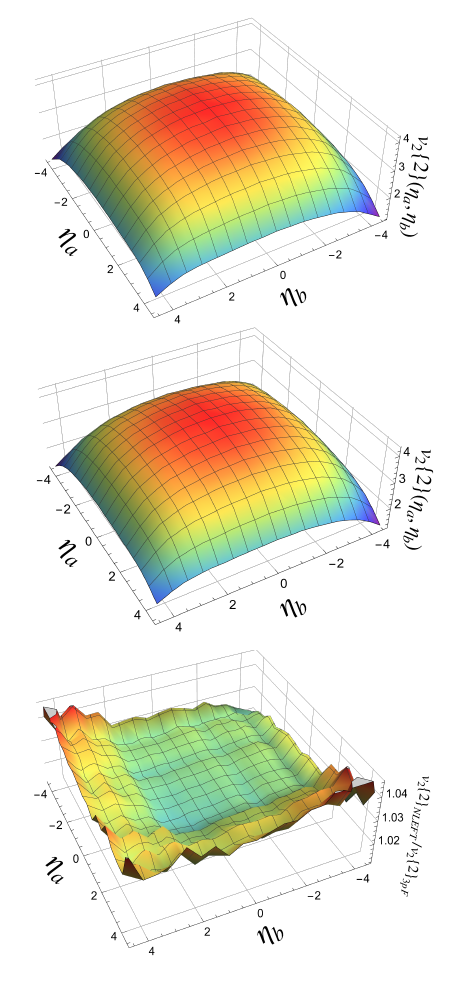}	
	\end{tabular}
	\begin{picture}(0,0)
	\put(-85,485){{\fontsize{10}{10}\selectfont \textcolor{black}{\rotatebox{15}{O+O, 3pF}}}}
	\put(-85,320){{\fontsize{10}{10}\selectfont \textcolor{black}{\rotatebox{15}{O+O, NLEFT}}}}
	\put(-85,155){{\fontsize{12}{12}\selectfont \textcolor{black}{\rotatebox{15}{Ratio}}}}
	\put(25,48){{\fontsize{12}{12}\selectfont \textcolor{black}{\rotatebox{20}{$0-5\%$}}}}
	\put(80,480){{\fontsize{10}{10}\selectfont \textcolor{black}{\rotatebox{-40}{$\times10^{-2}$}}}}
	\put(80,315){{\fontsize{10}{10}\selectfont \textcolor{black}{\rotatebox{-40}{$\times10^{-2}$}}}}
	\end{picture}		
	\caption{ Two-dimentional elliptic flow distribution  $v_2\{2\}(\eta_a,\eta_b)$  for the range of $0.2\leq p_T\leq3$ GeV is demonstrated for 3pF (a) and NLEFT (b) in $0-5\%$ centrality at $200$ GeV.  Since we find a similar pattern for the other two models (PGCM and VMC), we show the results of NLEFT in panel (b). To study the nucleon-nucleon correlations in the configurations computed by NLEFT for oxygen, the correlator $v_2\{2\}(\eta_a,\eta_b)_\text{NLEFT}$ is normalized by $v_2\{2\}_\text{3pF}$ in panel (c).} 
	\label{fig2}
\end{figure}
\section{Results}\label{sec3}
 In this section, we present the first findings for longitudinal decorrelations in O+O collisions at 200 GeV, highlighting the inherent nuclear characteristics as captured by three distinct models: NLEFT, PGCM, and VMC (Sec.\ref{sec3a}). We undertake a comparative analysis of O+O and d+Au collisions to investigate the influence of system size and nuclear configurations in Sec.\ref{sec3b}. Furthermore, we extend this methodology to examine O+O and Ne+Ne collisions within the framework of the NLEFT model, offering predictions for forthcoming LHC Runs.

\begin{figure}[t!]
	\begin{tabular}{c}
		\includegraphics[scale=0.4]{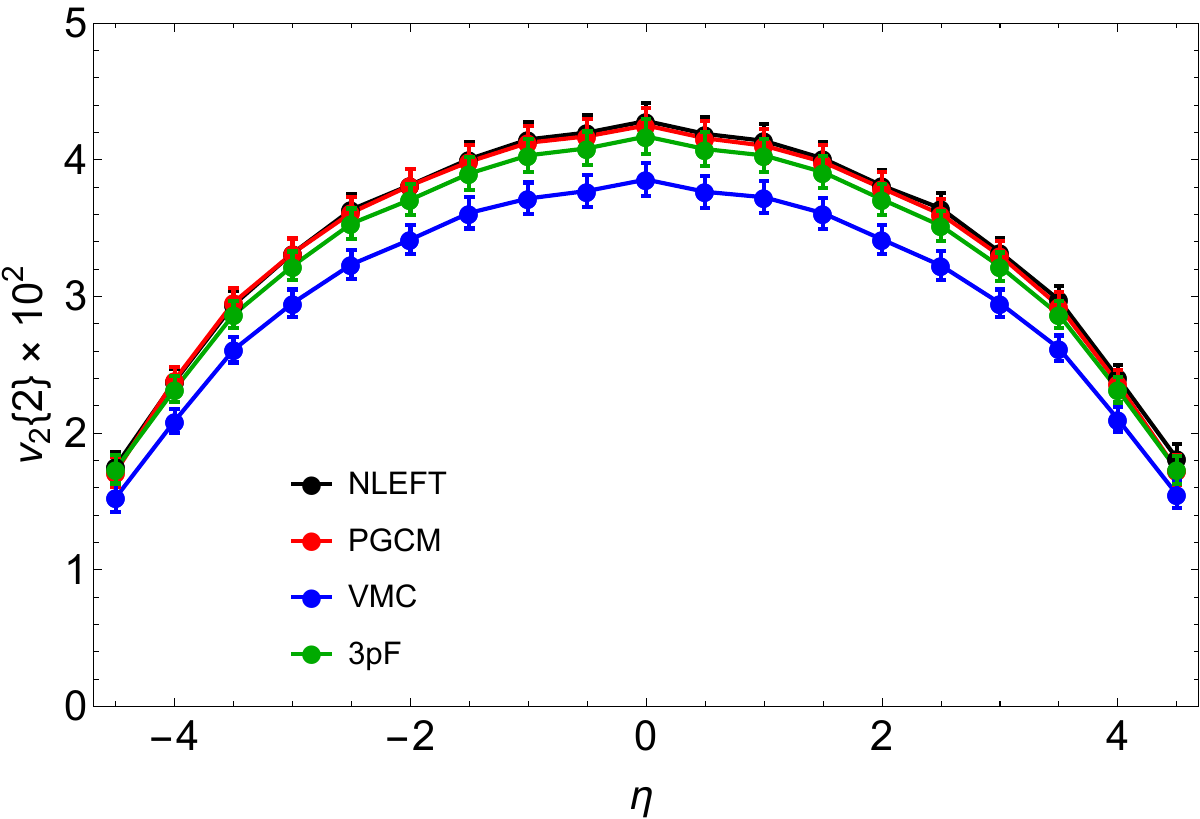}\\
		\includegraphics[scale=0.4]{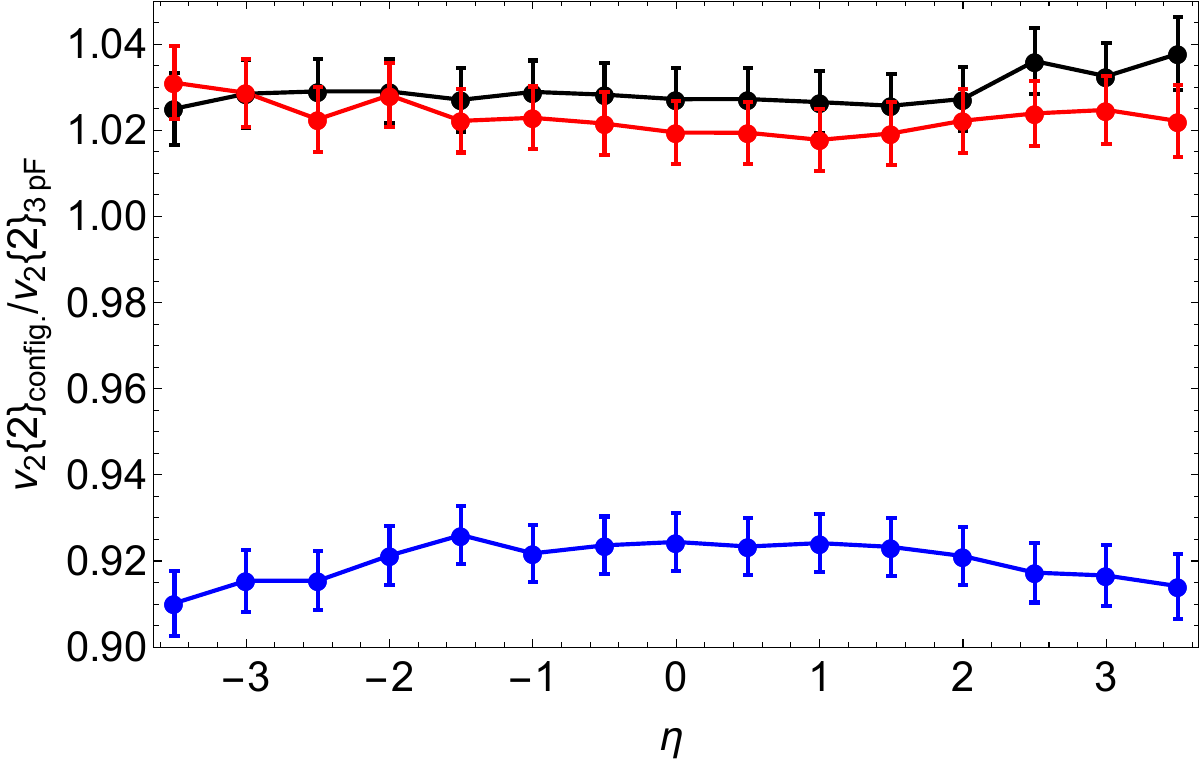}	
	\end{tabular}
	\begin{picture}(0,0)
	\put(-100,40){{\fontsize{9}{9}\selectfont \textcolor{black}{O+O @$200$ GeV}}}
	\put(-100,30){{\fontsize{9}{9}\selectfont \textcolor{black}{$0.2\leq p_T\leq3$ GeV}}}
	\put(-60,140){{\fontsize{9}{9}\selectfont \textcolor{black}{$0-5\%$}}}
	\put(-210,25){{\fontsize{9}{9}\selectfont \textcolor{black}{(a)}}}
	\put(-200,-90){{\fontsize{9}{9}\selectfont \textcolor{black}{(b)}}}
	\end{picture}		
	\caption{ (a) The diagonal terms of $v_2\{2\}(\eta_a,\eta_b)$ in Fig.\ref{fig2} are demonstrated for different \textit{ab-initio} models. (b) Also the ratio $v_2\{2\}_{\text{config}}/v_2\{2\}_{\text{3pF}}$ are computed for different oxygen structures. This ratio plays the role of a structure discriminator.} 
	\label{fig3}
\end{figure}
\begin{figure}[t!]
	\begin{tabular}{c}
		\begin{tabular}{c}
			\includegraphics[scale=0.4]{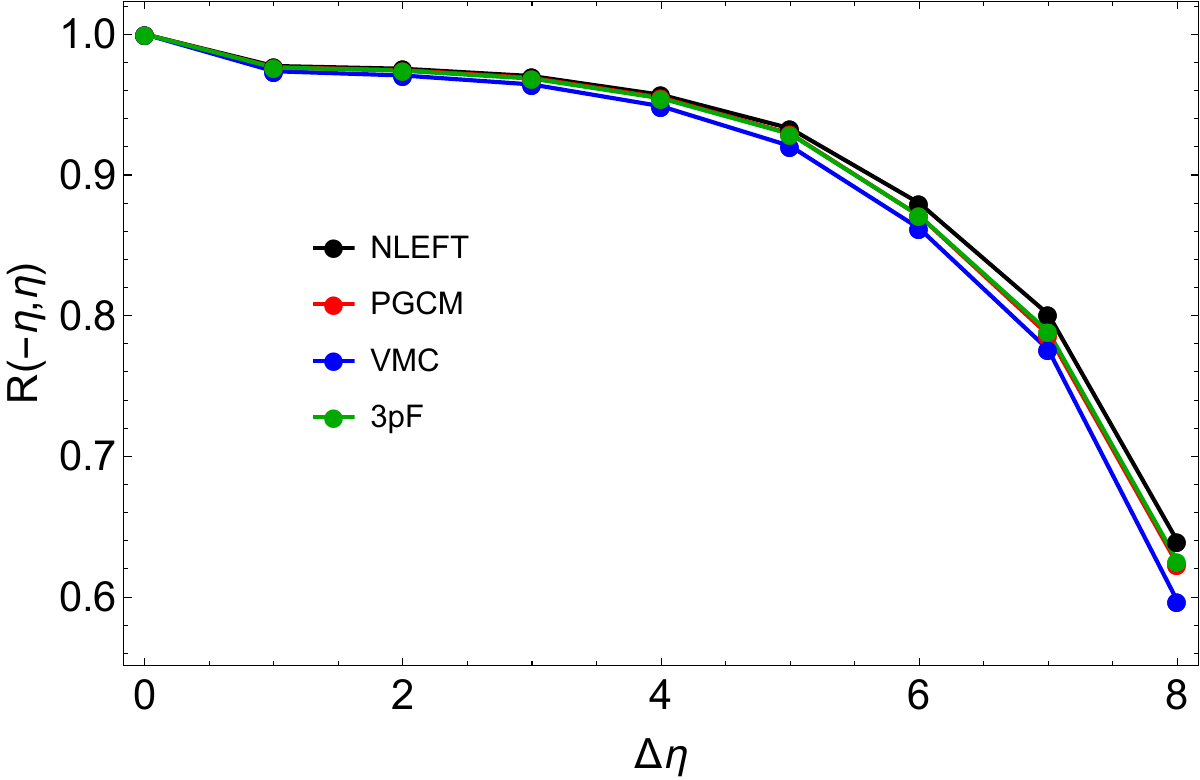}\\
			\includegraphics[scale=0.4]{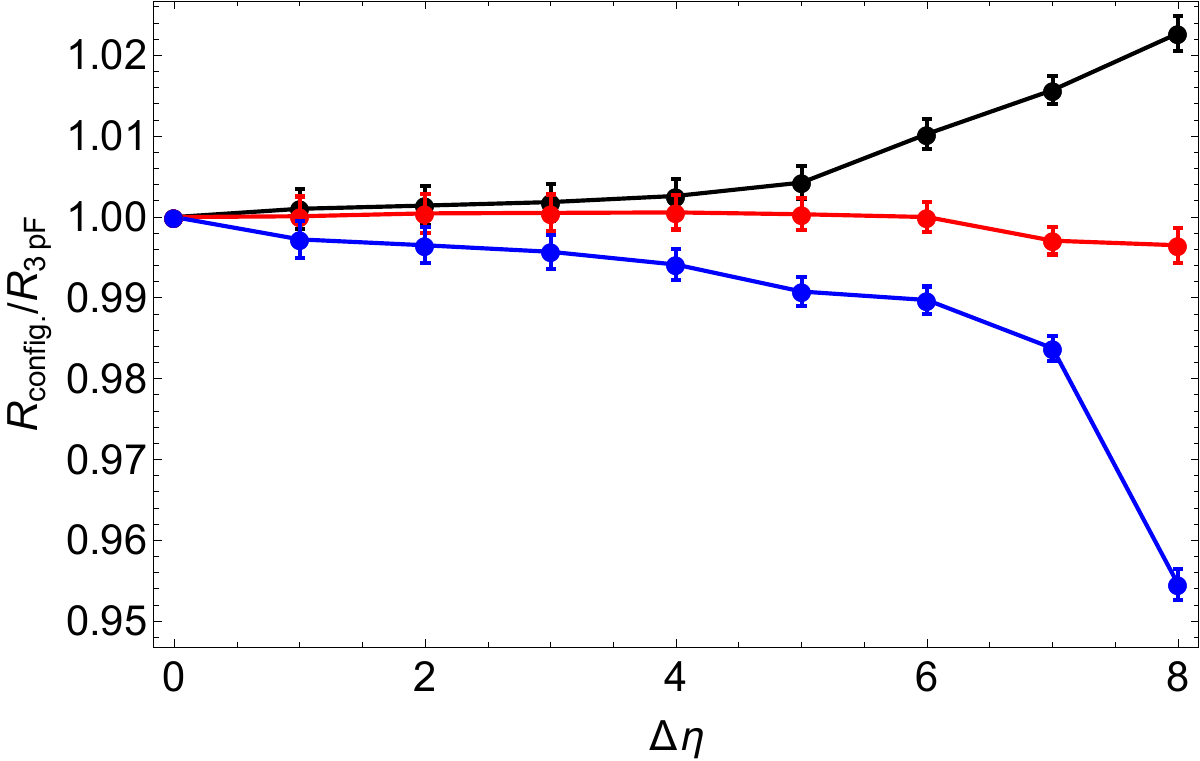}
		\end{tabular}	
	\end{tabular}
	\begin{picture}(0,0)
	\put(-190,50){{\fontsize{9}{9}\selectfont \textcolor{black}{O+O @$200$ GeV}}}
	\put(-190,40){{\fontsize{9}{9}\selectfont \textcolor{black}{$0.2\leq p_T\leq3$ GeV}}}
	\put(-170,-120){{\fontsize{9}{9}\selectfont \textcolor{black}{$0-5\%$}}}
	\put(-205,30){{\fontsize{9}{9}\selectfont \textcolor{black}{(a)}}}
	\put(-205,-120){{\fontsize{9}{9}\selectfont \textcolor{black}{(b)}}}
	\end{picture}
	\caption{Decorrelations of two pseudorapidity intervals at $\pm\eta$ in panel (a) and ratios (configs/3pF) to determine the different oxygen structures in panel (b) are presented. The results indicate that this correlator cannot identify PGCM from the spherical shape of $^{16}O$ (3pF).} 
	\label{fig4}
\end{figure}
\subsection{Discriminators of Oxygen Structures}\label{sec3a}
 Recently, several studies have been carried out to predict bulk observables of $^{16}$O, including elliptic flow and fluctuations in mean transverse momentum,  utilizing advanced \textit{ab-initio} models. \cite{Giacalone:2024luz,Huang:2023viw,Zhang:2024vkh,Giacalone:2024ixe,Mehrabpour:2025ogw,Zhao:2024feh}. These include: 1) NLEFT simulations utilizing a minimal pion-less EFT Hamiltonian \cite{Meissner:2014lgi,Elhatisari:2017eno}, 2) VMC simulations grounded on N$^2$LO chiral EFT Hamiltonian \cite{Lonardoni:2018nob}, and 3) PGCM calculations related to N$^3$LO chiral EFT Hamiltonian \cite{Frosini:2021fjf,Frosini:2021sxj,Frosini:2021ddm}. The different Hamiltonians and approximations employed result in varied tetrahedral-like clustering correlations \cite{Mehrabpour:2025ogw,YuanyuanWang:2024sgp}, which consequently lead to diverse predictions regarding observables of relativistic O+O collisions.  Although multiple studies have been conducted using these models, recent STAR results show that only the observables derived from VMC configurations show better alignment with RHIC data using the initial state observables \cite{Huang:2023viw}. 
Since the initial model results are not enough for comparison with the experimental data, we provide the results from fully hydro calculations in this work.	
In this section, we study different oxygen structures, NLEFT, PGCM and VMC, to give a prediction of longitudinal observables.  Additionally, we compare these models with the nucleon configurations of oxygen generated by the charge density 3pF parametrization of the nucleus:
\begin{equation}\label{q5}
	\rho(r)\propto \frac{1+w(r^2/R^2)}{1+e^{(r-R)/a_0}},
\end{equation}
where $R$ denotes the half-width radius, $w$ measures central density depletion, and $a_0$ characterizes the surface diffuseness. We considered $R=2.608$ fm, $a_0=0.513$ fm, and $w=-0.051$ to generate the charge density of the independent nucleon configurations for oxygen \cite{Zhang:2024vkh,ANGELI201369}. 
The information of flow decorrelations is encoded in the two-dimensional distribution of $v_2\{2\}(\eta_a,\eta_b)$ which is found by the 2-particle correlation (2PC) method \cite{Takahashi:2009na} in Eq.\ref{q1}. Fig.\ref{fig2}a and Fig.\ref{fig2}b display the distributions obtained by 3pF and NLEFT densities respectively in $0-5\%$ central O+O collisions at $200$ GeV center-of-mass energy.  As can be seen, the same effect is observed for these structures on flow decorrelations \footnote{We also checked PGCM and VMC, and then we found the same behavior.}. 
 In the case of light to intermediate-mass nuclei (with atomic mass number A around 20), traditional mean-field models become inadequate, and the significance of two-body nucleon-nucleon (NN) correlations increases substantially \cite{Cruz-Torres:2019fum}. Contemporary \textit{ab-initio} methods for tackling the nuclear many-body problem can now investigate deformed intermediate-mass nuclei and the development of clustering correlations within them from first principles \cite{Stroberg:2019bch,Otsuka:2022bcf}.
 To study NN correlations in high energy ion collisions, we normalize the flow decorrelations of Eq.\ref{q1} obtained for the different configurations to the results derived by the sampled nucleons concerning 3pF charge density distribution baseline of the independent nucleons \cite{Zhang:2024vkh}. The ratio $v_2\{2\}_{\text{NLEFT}}/v_2\{2\}_{\text{3pF}}$ highlights the impact of nontrivial nucleon-nucleon correlations in the longitudinal direction. Fig.\ref{fig2}c shows the deviation of this ratio from unity which implies the presence of $NN$ correlations in the NLEFT Hamiltonian. 

\begin{figure}[t!]
	\begin{tabular}{c}
		\begin{tabular}{c}
			\includegraphics[scale=0.4]{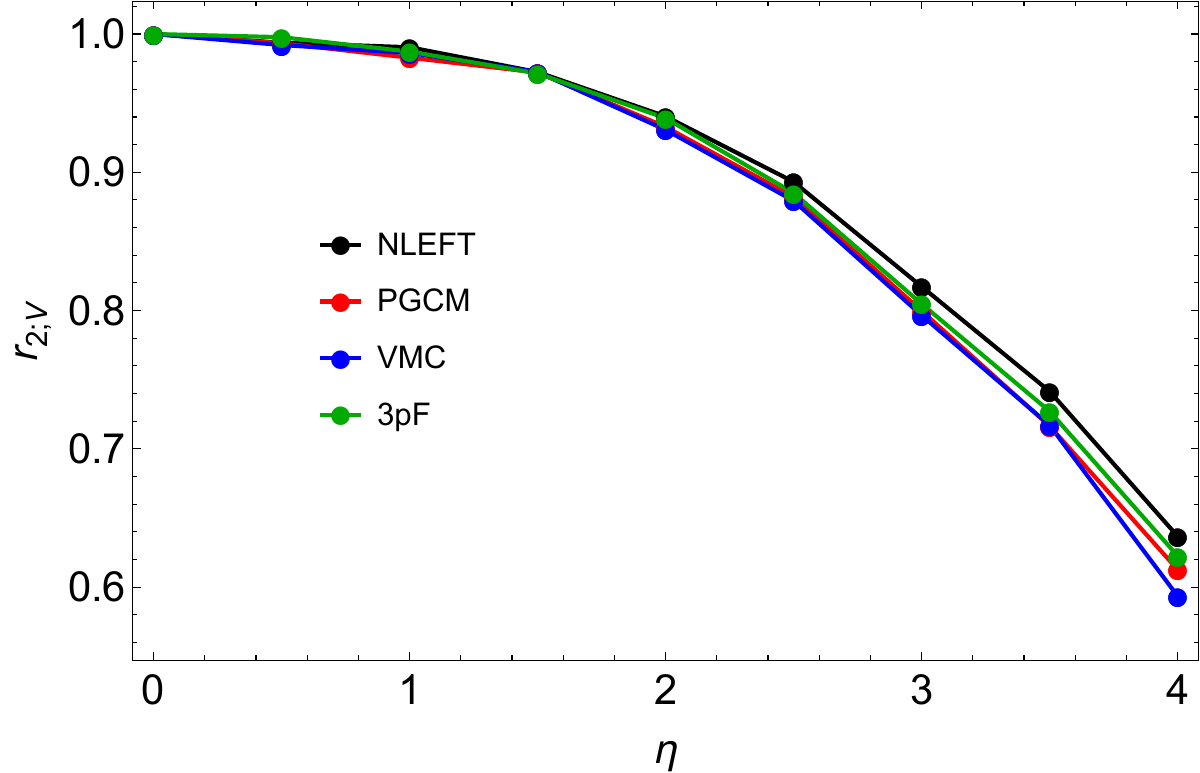}\\
			\includegraphics[scale=0.4]{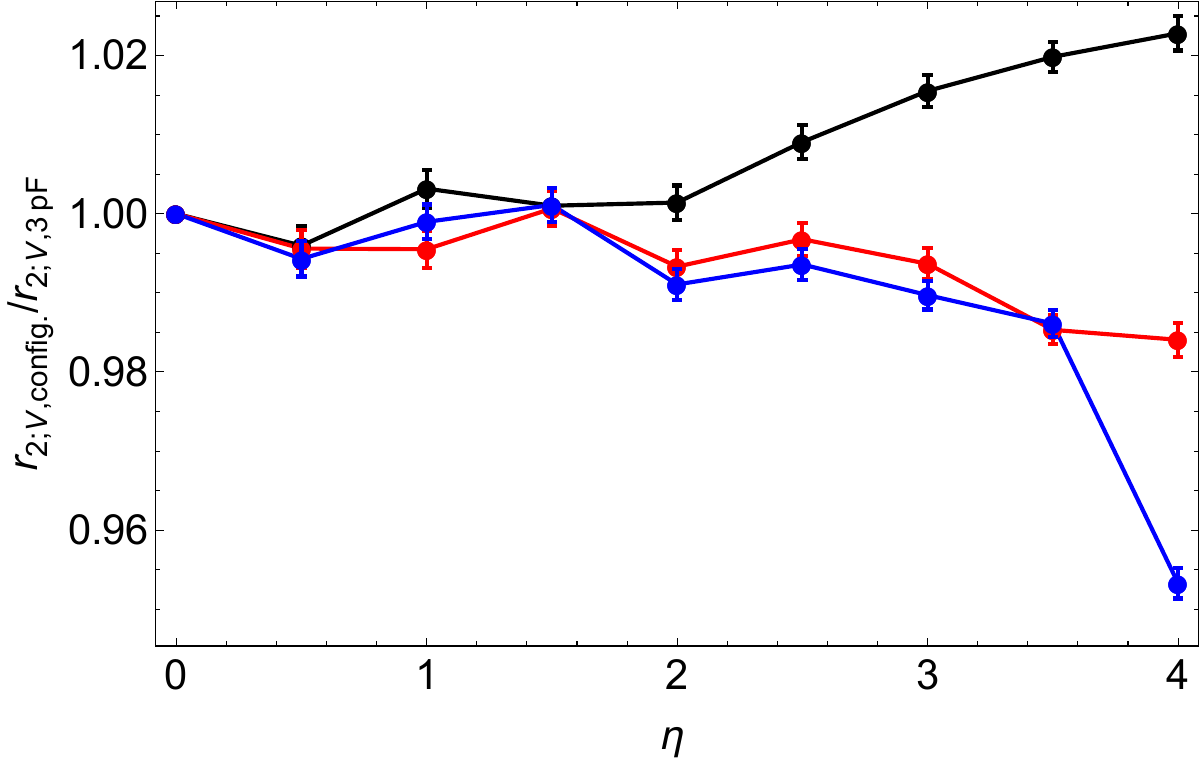}
		\end{tabular}	
	\end{tabular}
	\begin{picture}(0,0)
	\put(-150,50){{\fontsize{9}{9}\selectfont \textcolor{black}{O+O @$200$ GeV}}}
	\put(-150,40){{\fontsize{9}{9}\selectfont \textcolor{black}{$0.2\leq p_T\leq3$ GeV}}}
	\put(-45,140){{\fontsize{9}{9}\selectfont \textcolor{black}{$0-5\%$}}}
	\put(-200,30){{\fontsize{9}{9}\selectfont \textcolor{black}{(a)}}}
	\put(-200,-120){{\fontsize{9}{9}\selectfont \textcolor{black}{(b)}}}
	\put(-150,30){{\fontsize{9}{9}\selectfont \textcolor{black}{$4.75<\eta_{\text{ref}}<5.25$}}}
	\end{picture}
	\caption{Correlators $r_{2;V}$ (a) for NLEFT (black), PGCM (red), VMC (blue), and spherical shape of oxygen (green) are depicted. Also, discriminations of the modern models are presented in panel (b) using the ratio $r_{2;V,\text{configs.}}/r_{2;V,\text{3pF}}$. We note that the reference bin is located at $4.75<\eta_{\text{ref}}<5.25$.} 
	\label{fig5}
\end{figure}
To investigate the different oxygen structures, we
study diagonal ridges of $v_2\{2\}(\eta_a,\eta_b)_{\text{configs}}$ along
$\eta =(\eta_a+\eta_b)/2$ illustrated in Fig.\ref{fig3}a.  A same pattern is observed for amplitudes in positive and negative $\eta$ regions for all the models. Comparing of normalized configurations are also depicted in Fig.\ref{fig3}b. As can be seen, the nuclear structure contributions remain very strongly correlated across pseudorapidity. Ref.\cite{Zhang:2024vkh} showed that one can find the distinctions between VMC and other models concerning the ratio $v_2\{2\}(\eta)_{\text{config.}}/v_2\{2\}(\eta)_{\text{3pF}}$. Concerning this, Fig.\ref{fig3}b shows that, 
due to different encoding structure properties of models, the behavior of $v_2\{2\}(\eta)$ obtained by VMC configurations is differed up to $10\%$ from NLEFT and PGCM configurations. Moreover, the deviation from spherical structure is $\approx3\%$ for NLEFT (and PGCM) and $\approx9\%$ for VMC.
These results is qualitatively consistent with the calculations in Ref.\cite{Zhang:2024vkh}. 

\begin{figure}[t!]
	\begin{tabular}{c}
		\begin{tabular}{c}
			\includegraphics[scale=0.4]{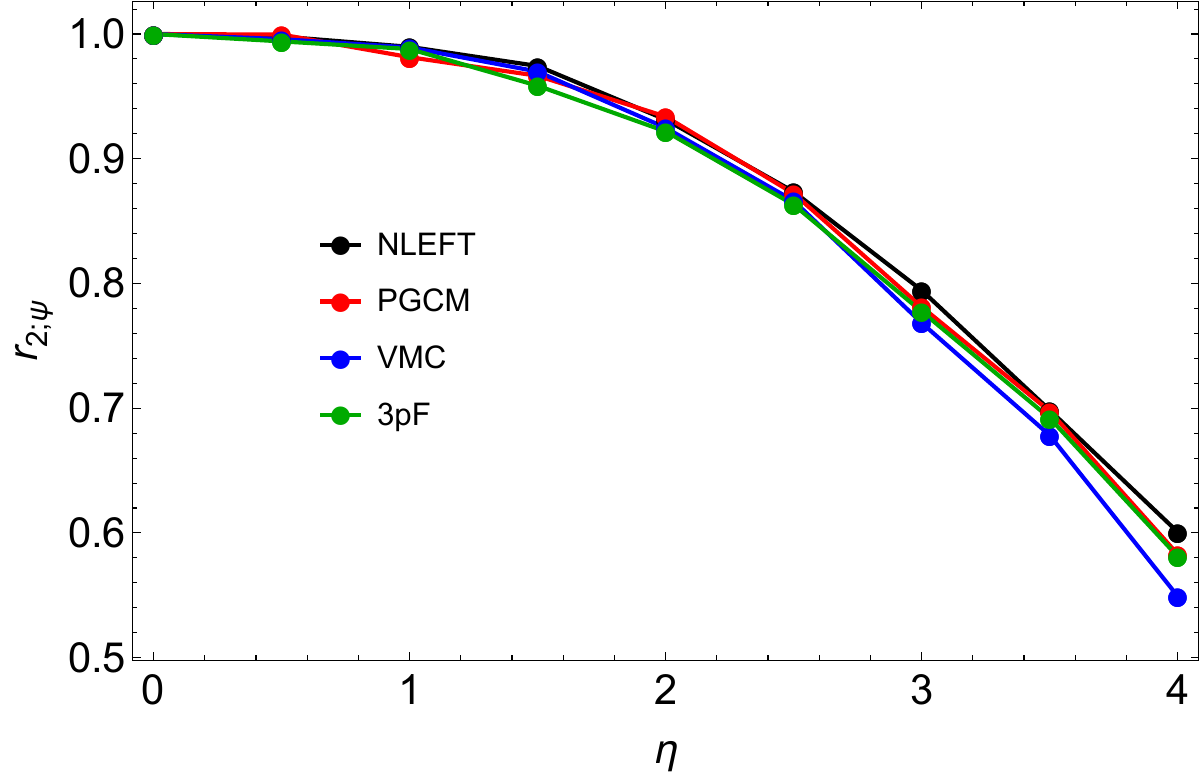}\\
			\includegraphics[scale=0.4]{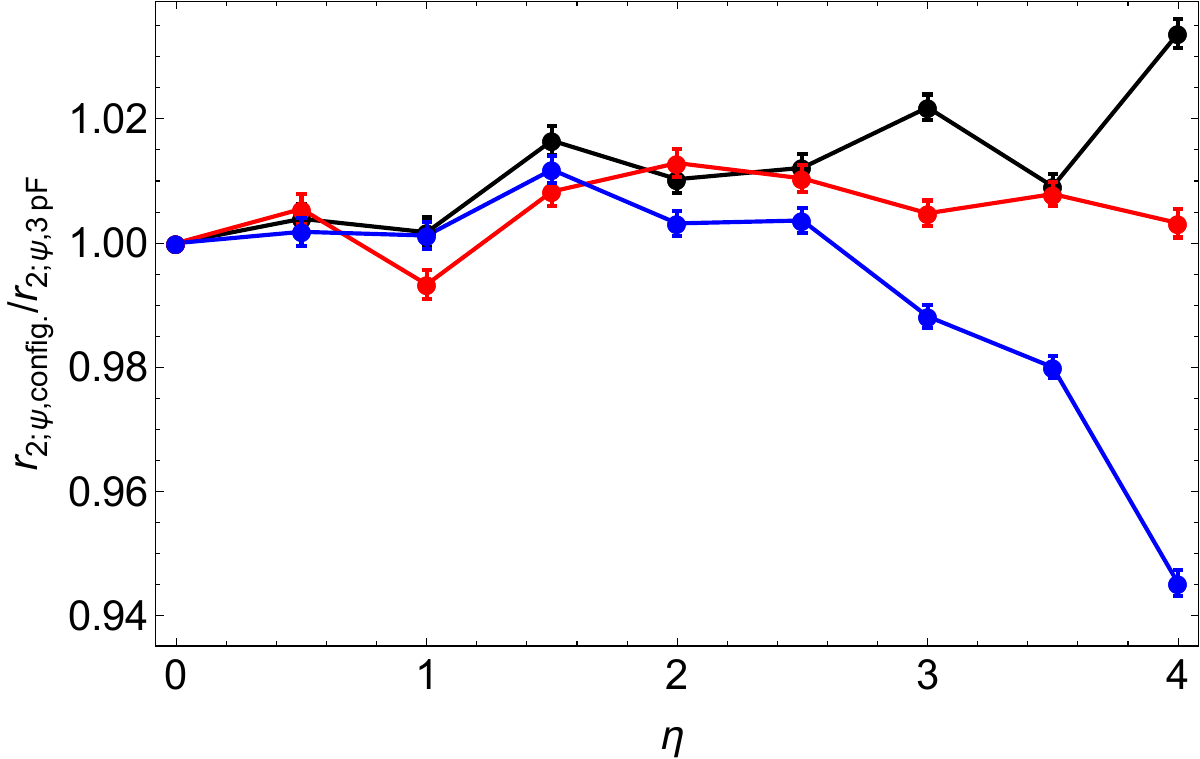}
		\end{tabular}	
	\end{tabular}
	\begin{picture}(0,0)
	\put(-150,50){{\fontsize{9}{9}\selectfont \textcolor{black}{O+O @$200$ GeV}}}
	\put(-150,40){{\fontsize{9}{9}\selectfont \textcolor{black}{$0.2\leq p_T\leq3$ GeV}}}
	\put(-45,140){{\fontsize{9}{9}\selectfont \textcolor{black}{$0-5\%$}}}
	\put(-200,30){{\fontsize{9}{9}\selectfont \textcolor{black}{(a)}}}
	\put(-200,-120){{\fontsize{9}{9}\selectfont \textcolor{black}{(b)}}}
	\put(-150,30){{\fontsize{9}{9}\selectfont \textcolor{black}{$4.75<\eta_{\text{ref}}<5.25$}}}
	\end{picture}
	\caption{Similar to Fig.\ref{fig5} for flow angle decorrelations $r_{2;\psi}$.} 
	\label{fig6}
\end{figure}
The degree of decorrelations manifests as deviation of the factorization $R(\eta_a,\eta_b)$ from unity. This quantity is strongly influenced by non-flow effect correlations \cite{CMS:2015xmx}.  Fig.\ref{fig4}a shows the results of the forward-backward decorrelations $R(-\eta,\eta)$ obtained from different configurations computed by stated models in O+O collisions at 0-5\% centrality. It can be seen for $\Delta\eta>3$ the decorrelations are started to increase significantly. However, the behaviors of decorrelations are the same for different models.  Concerning the ratio of $R(-\eta,\eta)_{\text{configs}}/R(-\eta,\eta)_{\text{3pF}}$, the splitting of different structures is appeared. Fig.\ref{fig4}b indicates that features of VMC are distinguishable from other two models such that they are $1\%$ for $\Delta\eta<4$, and $5\%$ with PGCM and $7\%$ with NLEFT in $\Delta\eta=8$.
Also, it indicates a separation between NLEFT and PGCM in $\Delta\eta>4$, although the maximum difference is up to $2\%$ in $\Delta\eta=8$.
\begin{figure}[t!]
	\begin{tabular}{c}
		\begin{tabular}{c}
			\hspace*{-1cm}\includegraphics[scale=0.41]{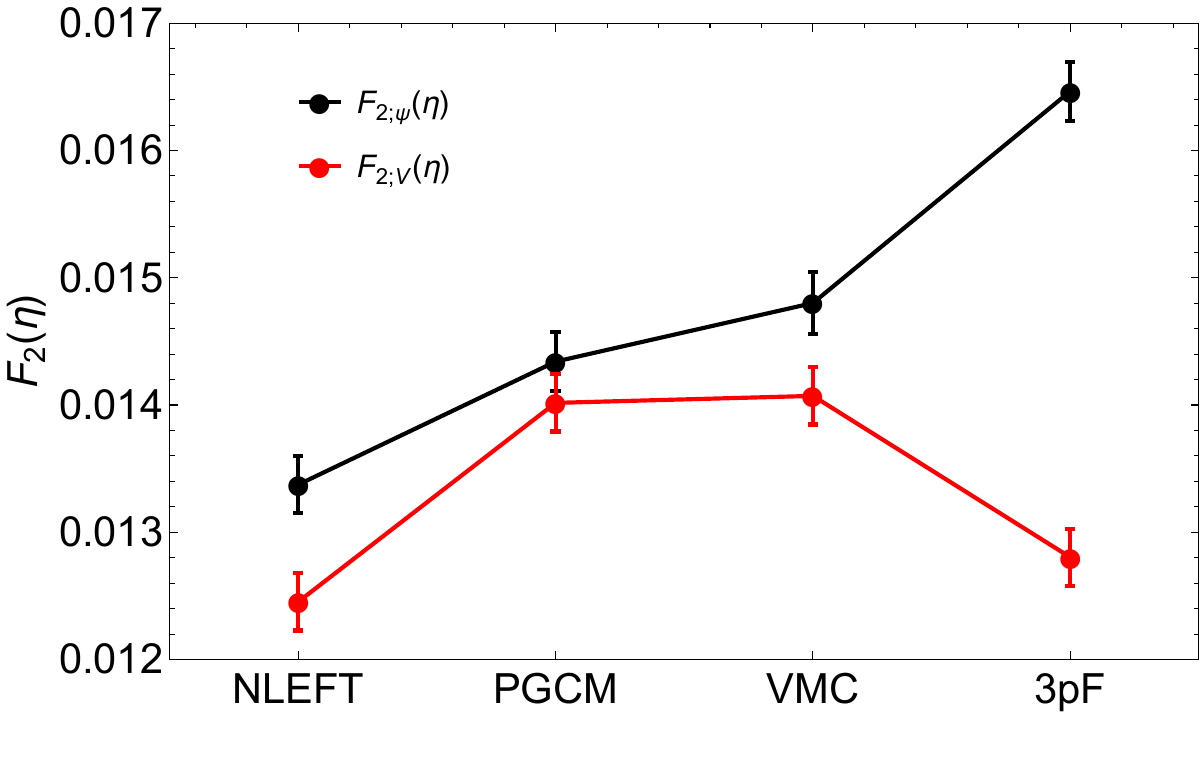}
		\end{tabular}	
	\end{tabular}
	\begin{picture}(0,0)
	\put(-140,-20){{\fontsize{9}{9}\selectfont \textcolor{black}{O+O @$200$ GeV}}}
	\put(-110,-40){{\fontsize{9}{9}\selectfont \textcolor{black}{$\eta\leq1.75$}}}
	\put(-140,-30){{\fontsize{9}{9}\selectfont \textcolor{black}{$0.2\leq p_T\leq3$ GeV}}}
	\put(-40,-45){{\fontsize{9}{9}\selectfont \textcolor{black}{$0-5\%$}}}
	\end{picture}
	\caption{The values of slop parameters $F_{2;\psi}$ (black) and $F_{2;V}$ (red) has been shown. As it is expected $F_{2;\psi}>F_{2;V}$ due to $r_{2;\psi}<r_{2;V}$. These values are found in $\eta\leq1.75$ where the correlators are almost linear.} 
	\label{fig7}
\end{figure}
Reducing non-flow contributions can be done using the correlators, $r_{2;V}$ and $r_{2;\psi}$, defined in Eqs.\ref{eq:Rv} and \ref{eq:Rc}. Figs.\ref{fig5} and \ref{fig6} present the results of the 3-pseudorapidity bin corralator of flow vector $V_2$ and orientation $\psi_2$, respectively. In our simulations, we take the reference bin as $4.75<\eta_{\text{ref}}<5.25$ far in the backward pseudorapidity region. We also symmetrize between the forward and backward pseudorapidities,
\begin{align*}
r_{2;V}(\eta)&=\frac{1}{2}\Big(\frac{v_2\{2\}^2(\eta,-\eta_{ref})}{v_2\{2\}^2(-\eta,-\eta_{ref})}+\frac{v_2\{2\}^2(-\eta,\eta_{ref})}{v_2\{2\}^2(\eta,\eta_{ref})}\Big),
\end{align*}
to increase statistics, which can be done for collisions of identical nuclei. The angle decorrelation can be symmetrized in a similar way. 
The results indicate same behavior for different structures in Fig.\ref{fig5}a and Fig.\ref{fig6}a. However, we differentiate between models using the ratio of $r_{2,\text{configs}}/r_{2,\text{3pF}}$ in panels (b). It can be observed that $r_{2;V}$ separates NLEFT from PGCM and VMC models in Fig.\ref{fig5}b. This separation would be stronger by choosing a bin close to $\eta_{ref}$ where the non-flow is increasing. The observable $r_{2;\psi}$ distinguishes VMC from NLEFT and PGCM  models, similar to $R(-\eta,\eta)$, as shown in Fig.\ref{fig6}b. In contrast, the correlator $r_{2;V}$ distinguishes between PGCM and 3pF, which cannot be captured from $r_{2;\psi}$ and $R(-\eta,\eta)$.
As has been shown in Fig.\ref{fig5}a and Fig.\ref{fig6}a, the longitudinal decorrelation functions are almost linear in the range of $\eta<2$, especially around mid-rapidity such that they can be parameterized as follows \cite{ATLAS:2017rij,Jia:2017kdq}:
\begin{align}
r_{2;V}(\eta)&=1-2\;\eta\;F_{2;V}(\eta),\label{qFv}\\
r_{2;\psi}(\eta)&=1-2\;\eta\;F_{2;\Psi}(\eta),
\end{align}  
where $F_{2,V}(\eta)$ and $F_{2,\Psi}(\eta)$ are called slope parameters. 
\begin{figure}[t!]
	\begin{tabular}{c}
		\begin{tabular}{c}
			\includegraphics[scale=0.44,angle=0]{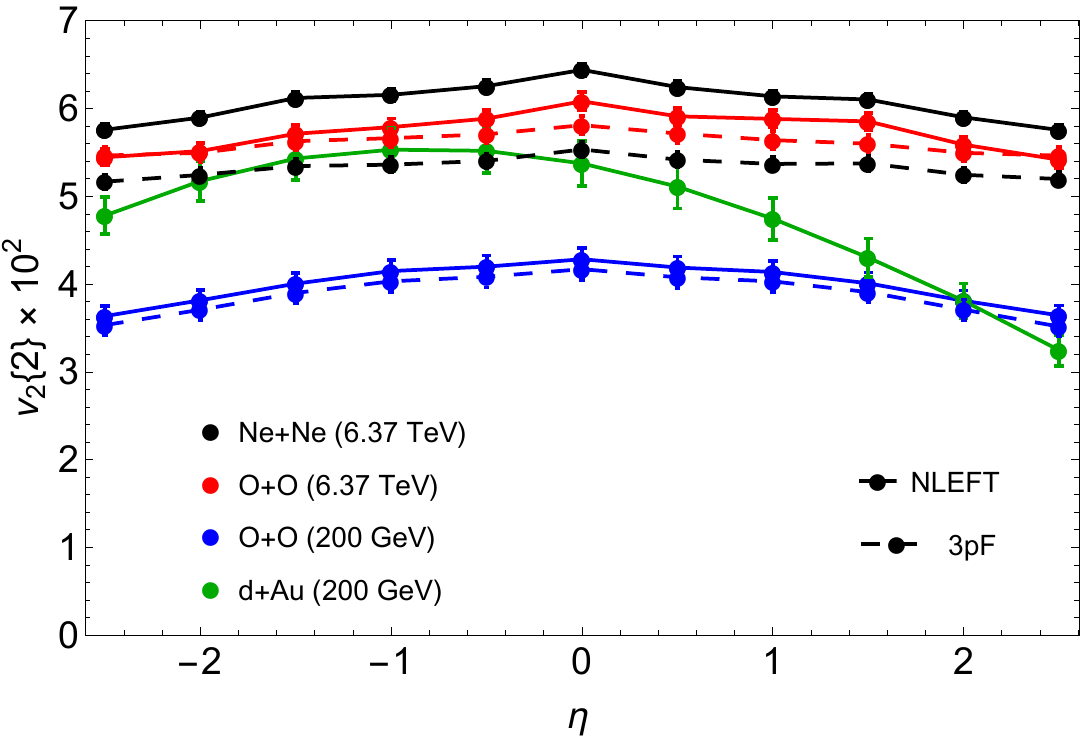}\\
			\includegraphics[scale=0.44]{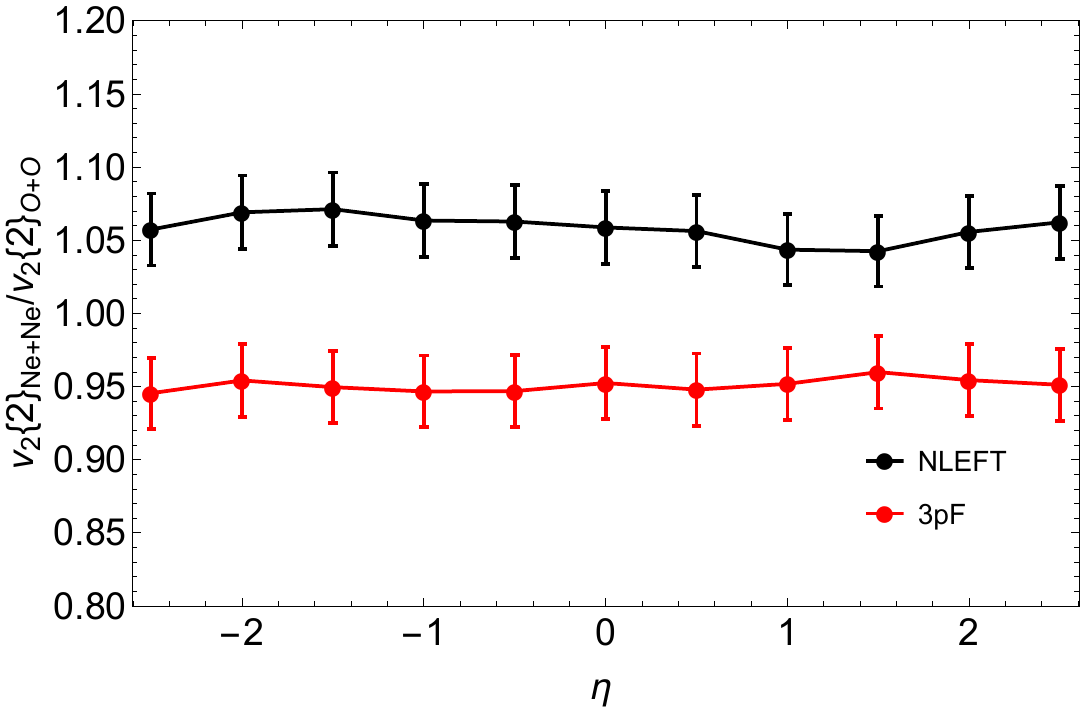}
		\end{tabular}	
	\end{tabular}
	\begin{picture}(0,0)
	\put(-215,30){{\fontsize{9}{9}\selectfont \textcolor{black}{(a)}}}
	\put(-205,-125){{\fontsize{9}{9}\selectfont \textcolor{black}{(b)}}}
	\put(-100,-20){{\fontsize{9}{9}\selectfont \textcolor{black}{$0-5\%$ 6.37 TeV}}}
	\end{picture}
	\caption{Diagonal terms of 2-dimensional flow distribution $v_2(\eta)$ are depicted in panel (a) for Ne+Ne (black), O+O (red and blue), and d+Au (green) collisions in $0-5\%$ centrality, similar to Fig.\ref{fig3}a. NLEFT configurations are considered for $^{20}$Ne and $^{16}$O.  To highlight the impact of $^{20}$Ne structure,  the ratio $v_2\{2\}_{\text{Ne+Ne}}/v_2\{2\}_{\text{O+O}}$ is illustrated in panel (b).} 
	\label{fig8}
\end{figure}
\begin{figure}[t!]
	\begin{tabular}{c}
		\begin{tabular}{c}
			\includegraphics[scale=0.44]{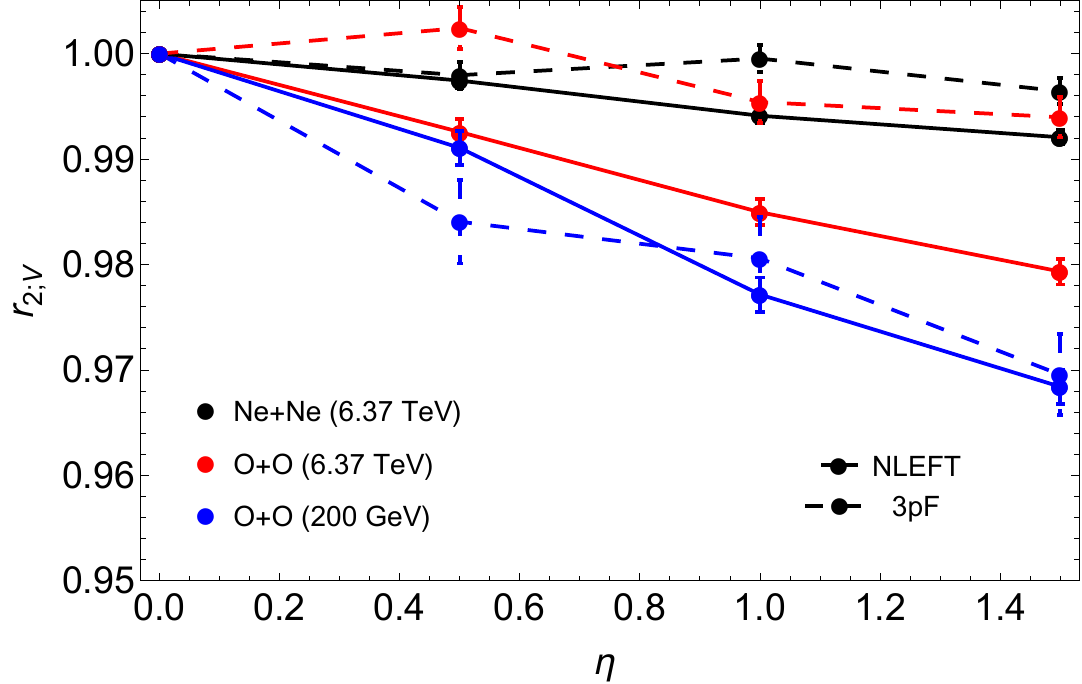}\\
			\includegraphics[scale=0.44]{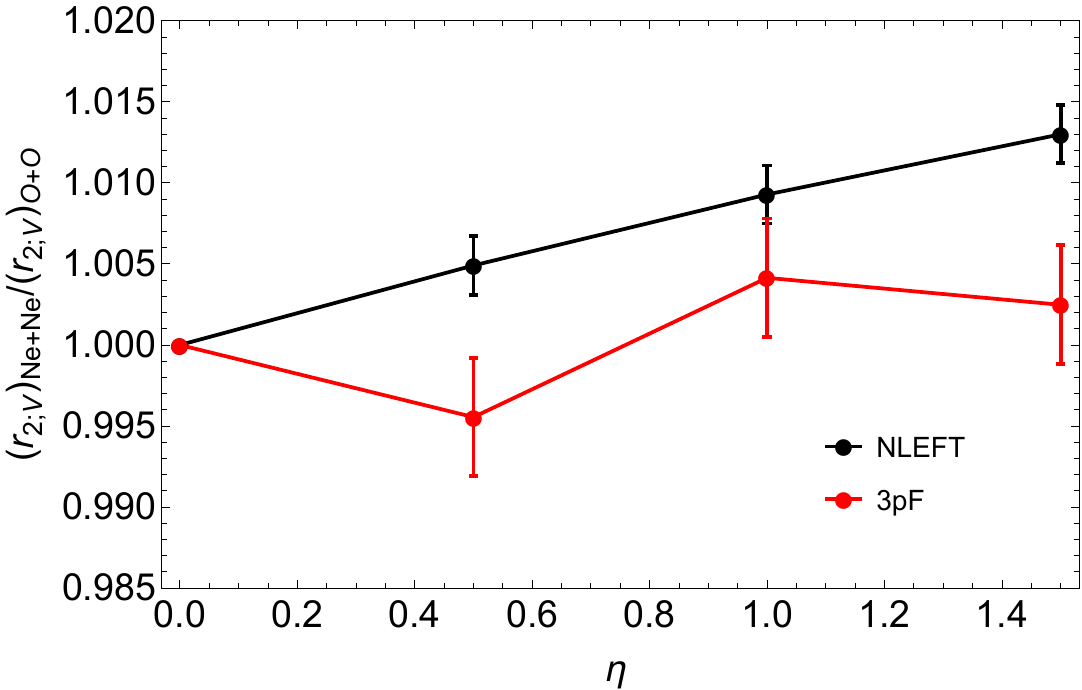}\\
			\includegraphics[scale=0.44]{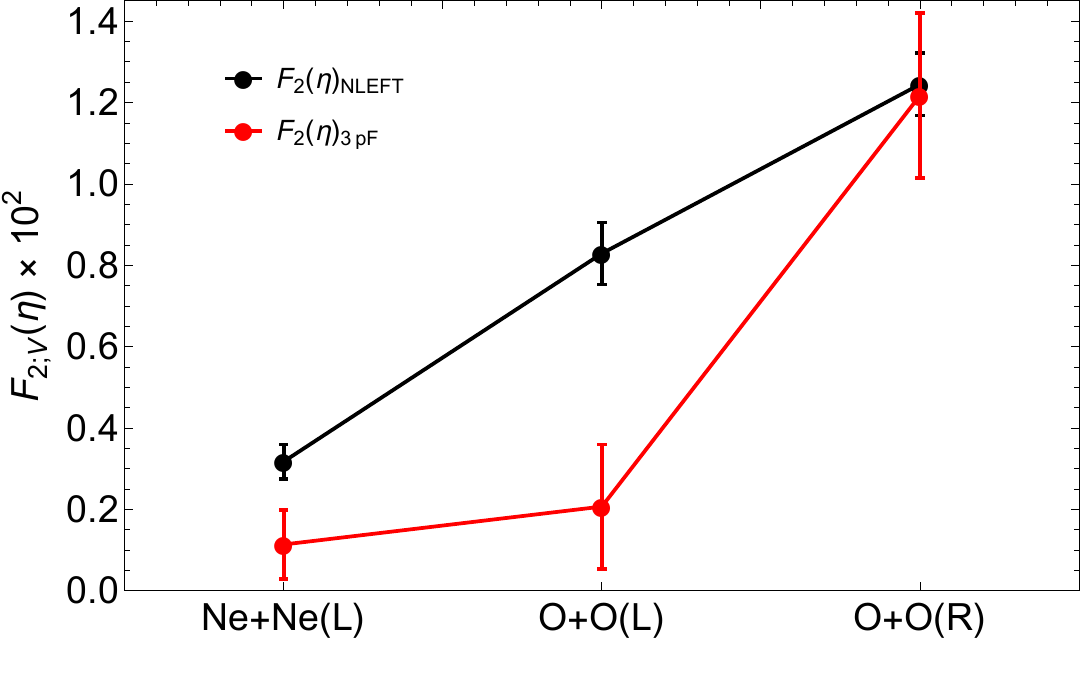}
		\end{tabular}	
	\end{tabular}
	\begin{picture}(0,0)
	\put(-30,111){{\fontsize{9}{9}\selectfont \textcolor{black}{(a)}}}
	\put(-30,-35){{\fontsize{9}{9}\selectfont \textcolor{black}{(b)}}}
	\put(-30,-190){{\fontsize{9}{9}\selectfont \textcolor{black}{(c)}}}
	\put(-80,-175){{\fontsize{9}{9}\selectfont \textcolor{black}{$-3\leq\eta\leq-3.5$}}}
	\put(-100,60){{\fontsize{9}{9}\selectfont \textcolor{black}{$0-5\%$ 6.37 TeV}}}
	\end{picture}
	\caption{(a) Correlator $r_{2,V}$ is obtained for symmetric collisions, Ne+Ne and O+O, at $0-5\%$ centrality. (b) To determine the rate of resembling of flow decorrelations the values of slop parameters $F_{2;V}$ (black) and $F_{2;\varepsilon_B}$ (red) have been shown.  To determine the impact of $^{20}$Ne structure,  the ratio $v_2\{2\}_{\text{Ne+Ne}}/v_2\{2\}_{\text{O+O}}$ for NLEFT (black) and 3pF (red) is illustrated in panel (b). Slope parameters for different collisions are displayed in panel (c). We note that ($L$) and ($R$) denote the collisions at LHC (6.37 TeV) and RHIC (200 GeV) energies.} 
	\label{fig9}
\end{figure}
\\ATLAS Collaboration suggested that these slope parameters can be measured by performing the $\eta$-weighted average for the deviation of correlation functions $r_{2;V}$ and $r_{2;\psi}$ from the unity \cite{Wu:2018cpc}:
\begin{equation}\label{eq:slop}
F_{2;}(\eta)= \frac{\sum_{i}\left\{1-r_{2;}(\eta_i)\right\}\eta_i}{2\sum_{i}\eta_i^2}.
\end{equation} 
 Here, we calculate these parameters to distinguish between different oxygen structures. These observables can be applied on experimental data to find the right nuclear structure and model. Figs.\ref{fig5} and \ref{fig6} indicate Eq.\ref{eq:slop} is a suitable discriminator for nuclear structures. As depicted in Fig.\ref{fig7}, discriminator $F_{2;V}$ is not able to differentiate between NLEFT and 3pF, as well as PGCM and VMC. However, the orientation discriminator presents distinct values for all discussed models quantitatively, and thus different nuclear structures can be quantified by employing this observable.

\begin{figure*}[t!]
	\begin{tabular}{c}
		\begin{tabular}{c}
			\includegraphics[scale=0.45,angle=0]{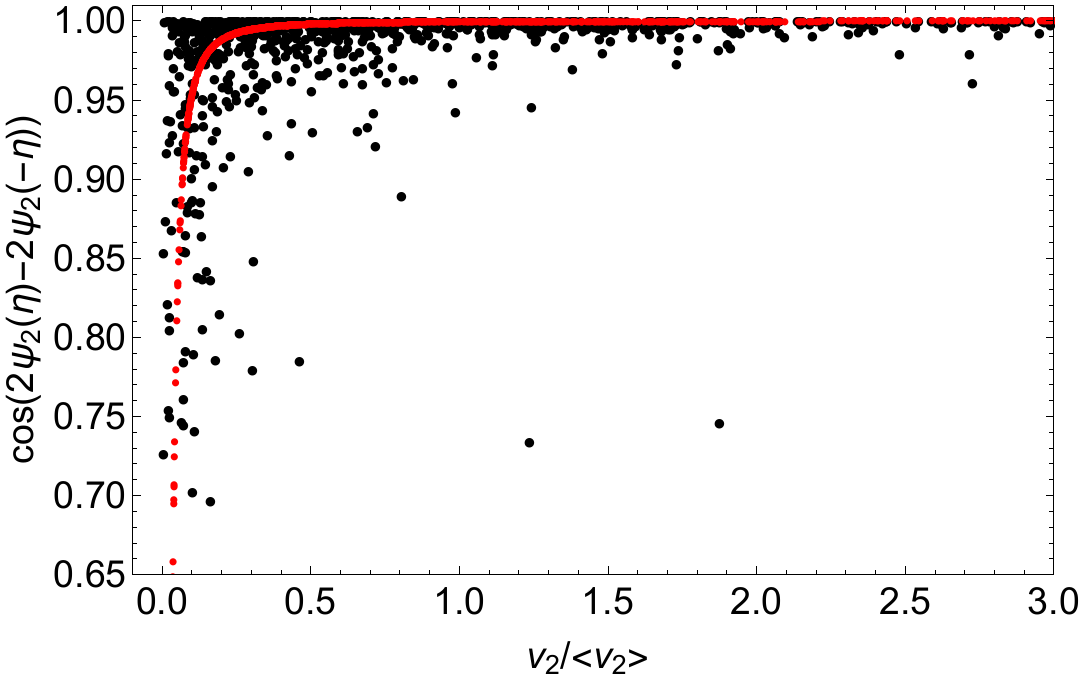}
			\includegraphics[scale=0.45,angle=0]{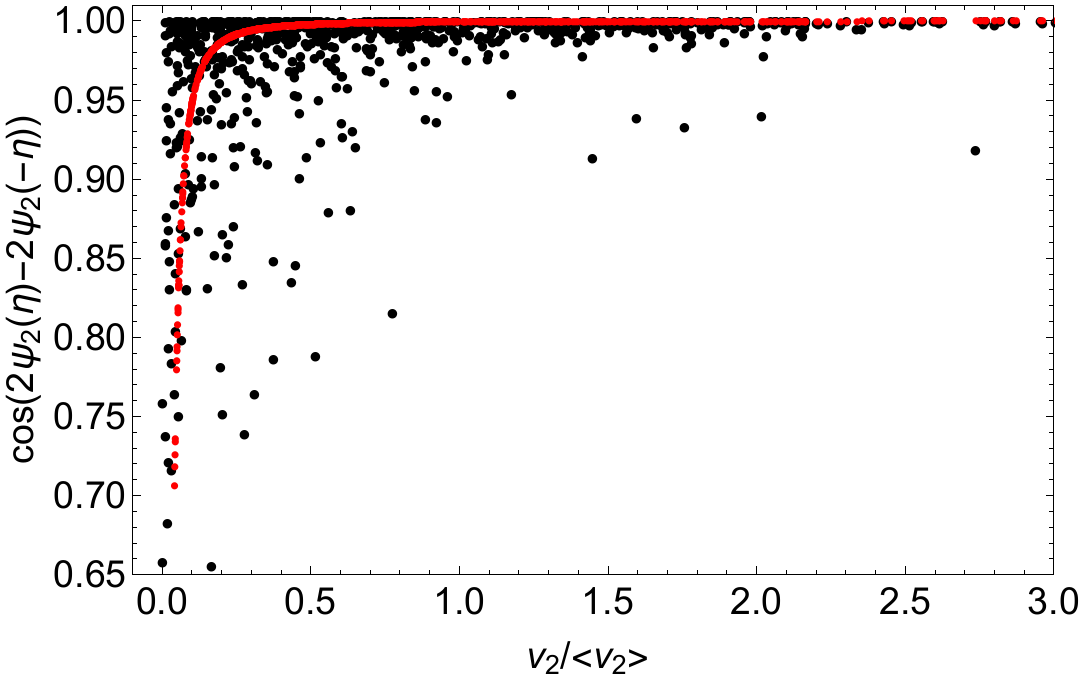}\\
			\includegraphics[scale=0.45,angle=0]{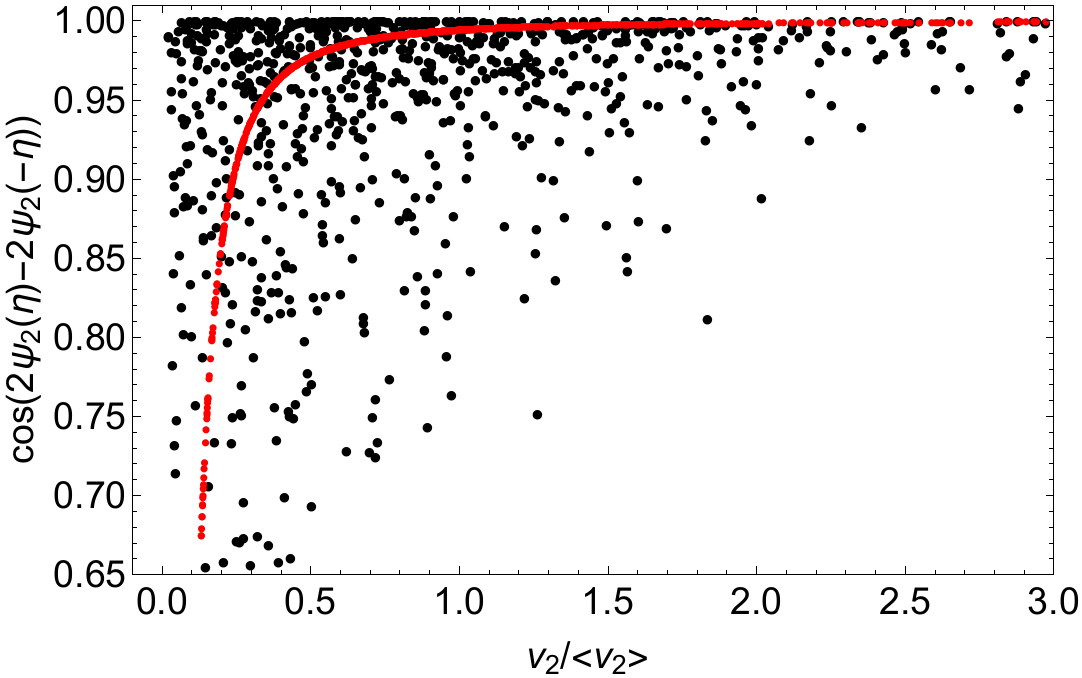}
			\includegraphics[scale=0.45,angle=0]{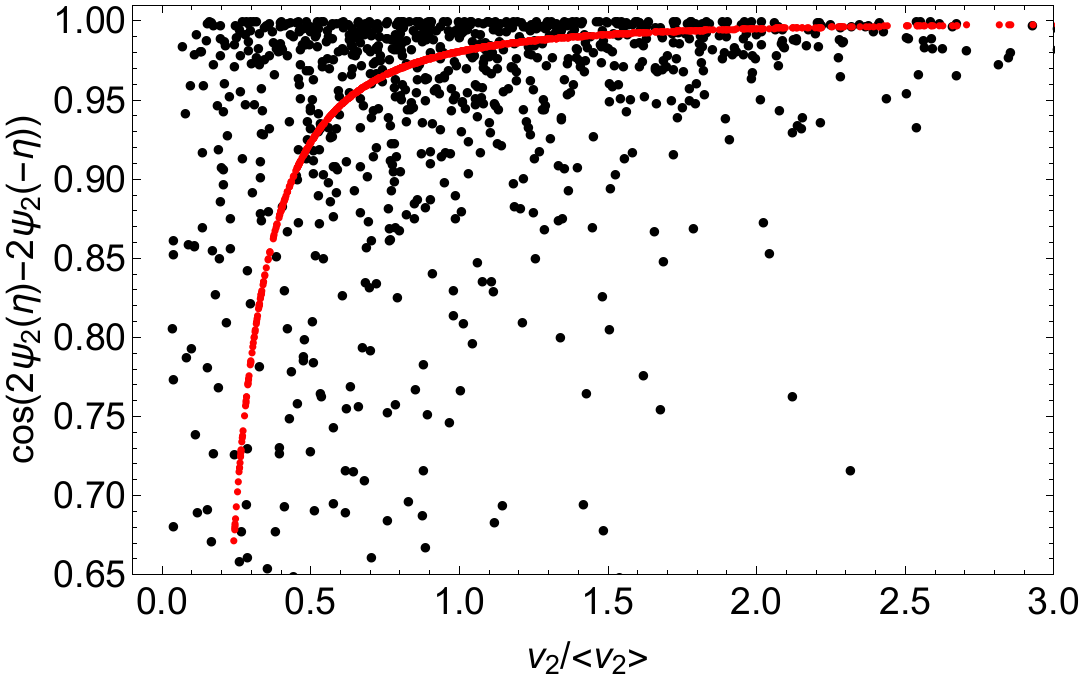}
		\end{tabular}	
	\end{tabular}
	\begin{picture}(0,0)
	\put(-270,35){{\fontsize{9}{9}\selectfont \textcolor{black}{(a)}}}
	\put(-30,35){{\fontsize{9}{9}\selectfont \textcolor{black}{(b)}}}
	\put(-30,-115){{\fontsize{9}{9}\selectfont \textcolor{black}{(d)}}}
	\put(-270,-115){{\fontsize{9}{9}\selectfont \textcolor{black}{(c)}}}
	\put(-350,35){{\fontsize{9}{9}\selectfont \textcolor{black}{Ne+Ne 6.37 TeV}}}
	\put(-100,35){{\fontsize{9}{9}\selectfont \textcolor{black}{O+O 6.37 TeV}}}
	\put(-350,-115){{\fontsize{9}{9}\selectfont \textcolor{black}{O+O 200 GeV}}}
	\put(-100,-115){{\fontsize{9}{9}\selectfont \textcolor{black}{d+Au 200 GeV}}}
	\put(-440,35){{\fontsize{9}{9}\selectfont \textcolor{black}{$0-5\%$}}}
	\end{picture}
	\caption{The scattered plot of $\cos(2\psi_2(\eta)-2\psi_2(-\eta))$ angle decorrelation versus the elliptic flow magnitude for the hydrodynamic model, for Ne+Ne (a), O+O(L) (b), O+O(R) (c) and d+Au (d) collisions at $0-5\%$ centrality. Here, we considered the NLEFT configurations for oxygen and neon. The pseudorapidity bin $\eta=2$ is considered here. The expected flow angle decorrelation as a function of the fixed value of flow is presented by red points.} 
	\label{fig10}
\end{figure*}
\subsection{ Light ions at RHIC and LHC Collisions}\label{sec3b}
As mentioned in Sec.\ref{sec2}, longitudinal initial fluctuations are encoded in decorrelation flow observables. These observables have been studied for different sizes and collision energies \cite{Zhang:2024bcb,Bozek:2018xzy,Wu:2021hkv}. Here, we do this study for d+Au (200 GeV), O+O (200 GeV and 6.37 TeV), and Ne+Ne (6.37 TeV) collisions. We should mention that here we generate Ne+Ne and O+O collisions concerning NLEFT model. Fig.\ref{fig8}a shows a comparison of elliptic flow at different $\eta$ bins obtained by 2PC method for various systems.  In this figure, we depicted Ne+Ne (black lines) and O+O (red lines) collisions at 6.37 GeV as well as  O+O (blue lines) and d+Au (green line) collisions at 200 GeV. Concerning $v_2\{2\}(\eta)$, identifying of different structures is accessible. As illustrated, we find:
\begin{align*}
	&v_2^{Ne+Ne} > v_2^{O+O(L)} > v_2^{d+Au}> v_2^{O+O(R)}\quad\text{for}\;\eta<2,\\
	&v_2^{Ne+Ne} > v_2^{O+O(L)} > v_2^{O+O(R)} > v_2^{d+Au}\quad\text{for}\;\eta>2,
\end{align*}
where  ($L$) and ($R$) indicate the $O+O$ collisions at LHC (6.37 TeV) and RHIC (200 GeV) energies, respectively. The results indicate the slope of $v_2\{2\}(\eta)$ is large in RHIC energies far from mid-rapidity. Moreover, the asymmetry of d+Au collision is manifested in this observable \cite{Shen:2016zpp}.  

 For completeness we repeated our calculations using configurations of $^{16}$O and $^{20}$Ne derived from a 3pF density distribution. The results are presented by dash lines in Fig.\ref{fig8}a. We use the mentioned parameters for oxygen in Sec.\ref{sec3a} as well as we parameterize Eq.\ref{q5} by $R_0=2.791$ fm, $a_0=0.698$ fm, $w=-0.168$ for $^{20}$Ne \cite{ANGELI201369}. The same trend is observed for O+O at RHIC energy for NLEFT and 3pF configurations. The slope of $v_2\{2\}(\eta)$ obtained by 3pF becomes smaller for light ion collisions at LHC energy in comparison with the NLEFT. Also, the tendency in Fig.\ref{fig8} is in a good consistency with the results of Ref.\cite{Giacalone:2024ixe} for these ions such that:
$$v_{2,\text{NLEFT}}^{Ne+Ne} > v_{2,\text{NLEFT}}^{O+O} > v_{2,\text{3pF}}^{O+O}> v_{2,\text{3pF}}^{Ne+Ne}\quad\text{at 6.37 TeV}.$$
To find the impact of Ne structure. we display the ratio $v_2\{2\}_{Ne+Ne}/v_2\{2\}_{O+O}$ in Fig.\ref{fig8}b. As can be seen, the results of panel (a) are confirmed in panel (b). The results indicate this ratio is approximately constant in the range of $-2.5< \eta<2.5$ and the difference is observed around 5\%.

 To avoid asymmetric collision effects and decrease non-flow contributions, we study the correlator $r_{2;V}$ in Eq.\ref{eq:Rv} for the symmetric collisions. The results in $0-5\%$ centrality are displayed in Fig.\ref{fig9}a  for NLEFT and 3pF configurations. It worths mentioning that the correlators are shown up to $\eta=1.75$ where they are almost linear. This enables us to investigate slope parameter for different collisions.  Also, we consider the reference bin is located $-3\leq\eta\leq-3.5$. As expected, the trends of Fig.\ref{fig8} also exist for NLEFT (solid lines) in this figure such that we have $r_{2}^{Ne+Ne}>r_{2}^{O+O(L)}>r_{2}^{O+O(R)}$.  This means that we have a stronger flow decorrelation for oxygen. The results of 3pF (dash lines) indicate $r_{2}^{Ne+Ne}\approx r_{2}^{O+O(L)}>r_{2}^{O+O(R)}$. Additionally, Fig.\ref{fig9}a displays $r_{2,\text{3pF}} > r_{2,\text{NLEFT}}$ which indicates a weak flow decorrelation for 3pF in comparison with NLEFT, and it highlights the effect of this structures in forward-backward flow correlations. The ratio $r_{2;V}^{Ne+Ne}/r_{2;V}^{O+O}$ is also studied for the collisions at 6.37 TeV in panel (b). It shows similar results for the 3pF configurations of oxygen and neon. However, the difference for NLEFT structures is obtained up to 1.5\%. 

 Slope parameters $F_{2;V}$ of Eq.\ref{eq:slop} for Ne+Ne and O+O collisions at 6.37 TeV as well as O+O collisions at 200 GeV are depicted in Fig.\ref{fig9}c. The results for NLEFT and 3pF are illustrated by black and red lines, respectively. As shown in the panel (c), the slope parameter can distinguish different collisions; however, similar values are obtained for Ne+Ne and O+O(L) in the 3pF configurations. Moreover, unlike RHIC energy, different structures, NLEFT and 3pF, can be identified in high energy at LHC by employing $F_{2;V}$.

Estimation of elliptic flow angle decorrelation between two bins centered at $\eta$ and $-\eta$ is defined \cite{Bozek:2017qir,Bozek:2023dwp}:
\begin{equation}
	\left\langle \cos\left(2 \psi_2(\eta)-2\psi_2(-\eta)\right) \right\rangle =\left\langle \frac{V_2(\eta)V_2^\star(-\eta)}{|V_n(\eta)||V_n(-\eta)|}\right\rangle \ .
	\label{eq:angle}
\end{equation} 
 Unlike $V_{2\Delta}(\eta,-\eta)$, flow angle decorrelation defined in Eq.\ref{eq:angle} cannot be measured experimentally. However, since the flow angle decorrelation is largest if the magnitude of overall flow is small, scatter plot of $\cos(2\Delta\psi_2)$ and overall flow magnitude gives us a useful information about higher moments of the overall flow magnitude \cite{Bozek:2017qir}. Moreover, it can be shown that the decorrelations is smaller in higher moments. In this way, we study how this inverse relation between $\cos(2\Delta\psi)$ and $v_2$ would be for different system collisions. The scattered plots are displayed in Fig.\ref{fig10}. Different collisions can be understood from the distributions in this figure such that flow angle fluctuates for d+Au in a wider range than other collisions. The most flow angle fluctuations in Ne+Ne and O+O collisions occur while $v_2$ is close to zero. This pattern is unlike what happens in large systems \cite{Bozek:2023dwp}. 
Ref.\cite{Bozek:2023dwp} shows that an increase in $\cos(2\Delta\psi)$ leads to a decrease in overall flow magnitude $v_2$ in the context of a random model. The anticorrelation of the flow angle decorrelation $\Delta\psi_2$ with $v_2$ is depicted as the red points in Fig.\ref{fig10}. We note that the relation between flow angle decorrelation and overall flow magnitude tells us more details about the collision energies as well as type of systems and it can be concluded using this anticorrelation.

\section{Conclusion}\label{conclusion}
This work presents a systematic investigation of longitudinal flow decorrelations as a probe of nuclear structure in light-ion collisions across RHIC and LHC energies. By combining ab-initio nuclear models (VMC, NLEFT, PGCM) with spherical 3pF densities, we establish a comprehensive study to quantify how initial-state nucleon configurations manifest in final-state collective dynamics. Our analysis spans symmetric (O+O, Ne+Ne) and asymmetric (d+Au) systems, leveraging longitudinal flow correlators to discriminate structure-driven effects in light ion collisions.

We first examined nucleon correlations from NLEFT, PGCM, and VMC models of oxygen at 200 GeV, relevant to RHIC data. We found that the ratio of flow vector and angle decorrelations from these models to the 3pF can differentiate various Hamiltonian structures, with the slope parameter also serving as a discriminator among theories. To assess the impact of oxygen shape, we analyzed d+Au collisions at the same energy. Flow vector correlations revealed asymmetry effects in these collisions, showing a more elliptical shape for the deuteron and larger flow correlations compared to O+O collisions. However, flow angle correlations exhibited a wider range of fluctuations in d+Au than in other collision types.

To study the effect of collision energies, we compared O+O collisions at 200 GeV and 6.37 TeV. The difference in energies affected not only the magnitude of $v_2$ but also resulted in smaller flow decorrelations at 6.37 TeV, with the slope parameter confirming this difference. Higher energies were associated with reduced flow angle fluctuations. We also examined the $^{20}$Ne structure on nucleon configurations from NLEFT and 3pF using longitudinal flow vector and angle correlations. NLEFT results for Ne+Ne collisions at 6.37 TeV showed a larger $v_2$ compared to O+O, while 3pF yielded a smaller  $v_2$ . The ratio Ne+Ne/O+O revealed a consistent 6\% difference for both NLEFT and 3pF configurations. Flow decorrelations in Ne+Ne collisions were smaller than in O+O for NLEFT configurations, while 3pF showed minimal decorrelations in both collision types. Flow angle fluctuations were similar for Ne+Ne and O+O collisions at 6.37 TeV.

Our results demonstrate that flow decorrelations serve as a sensitive tool for nuclear structure, with distinct signatures emerging across collision systems and energies. The identified discriminators provide a blueprint for future studies of deformation in lower to medium mass nuclei. The collected RHIC data and upcoming LHC oxygen runs can critically test these observations, while Bayesian methods could further refine structure-to-observable mappings. This work establishes a foundation for precision high-energy probes of nuclear geometry.

\section*{Acknowledgments}
We thank Huichao Song, Jiangyong Jia for insightful discussions. We are grateful to the participants of the 4th international workshop on QCD collectivity at the smallest scales for useful conversations. This work is supported in part by the National Natural Science Foundation of China under Grant No. 12247107.

\appendix
\begin{figure}[t!]
	\begin{tabular}{c}
		\begin{tabular}{c}
			\includegraphics[scale=0.41,angle=0]{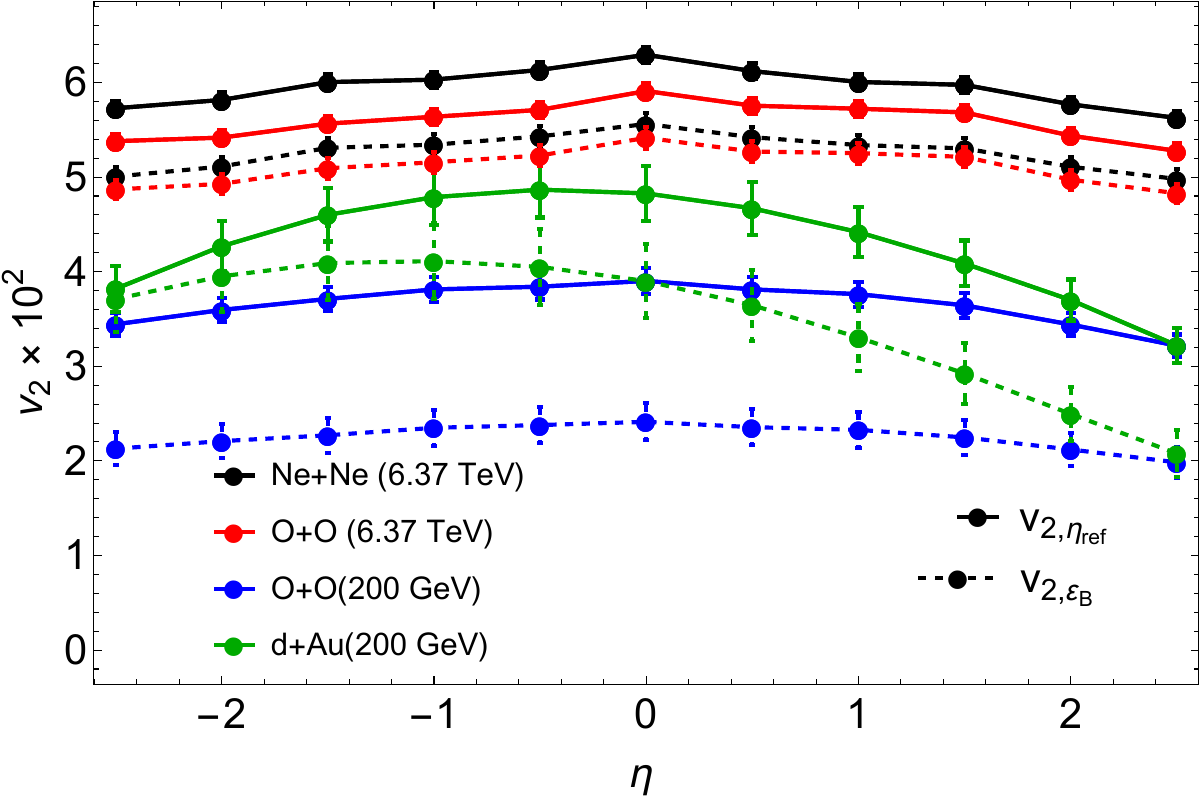}\\
			\includegraphics[scale=0.41]{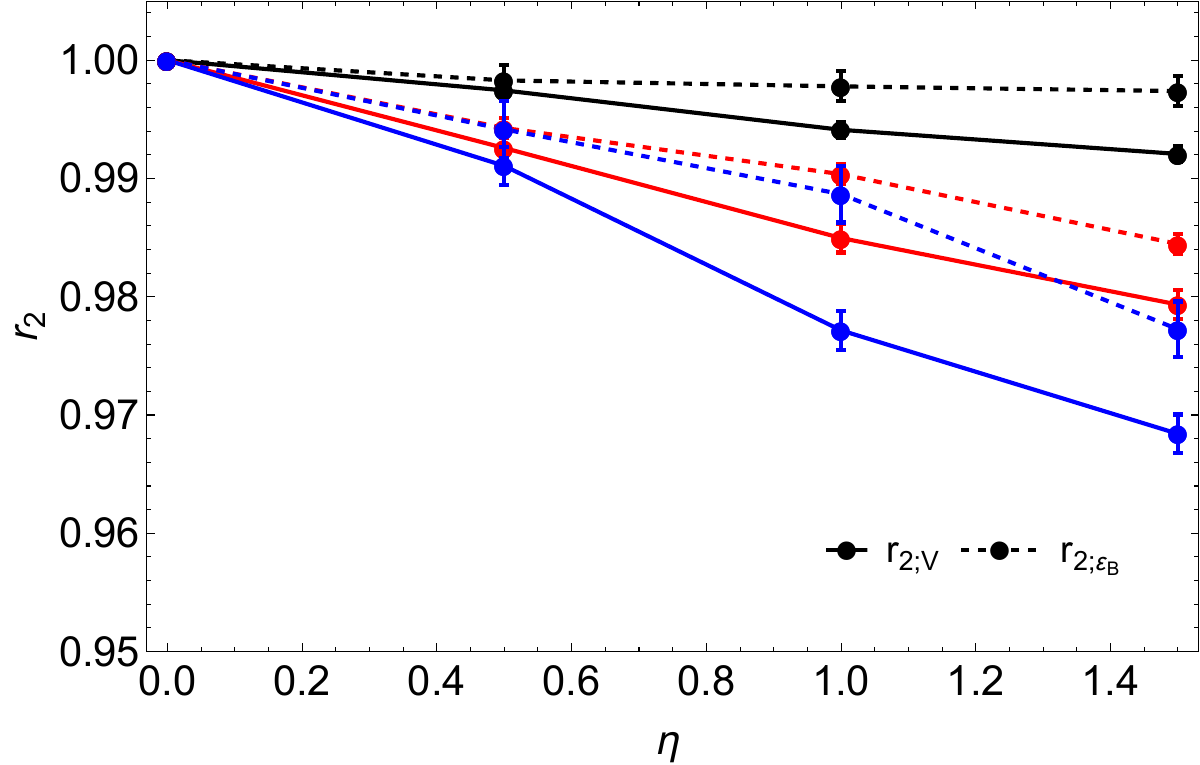}\\
			\includegraphics[scale=0.41]{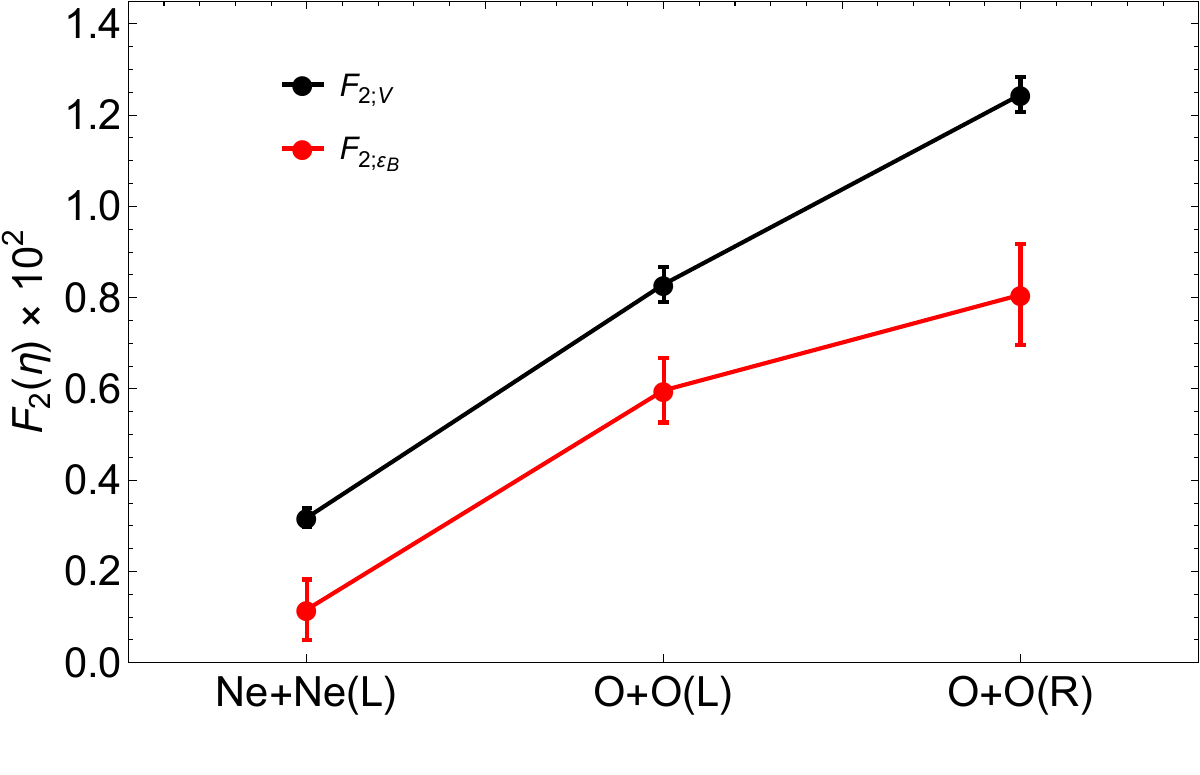}
		\end{tabular}	
	\end{tabular}
	\begin{picture}(0,0)
	\put(-95,341){{\fontsize{9}{9}\selectfont \textcolor{black}{(a)}}}
	\put(-85,190){{\fontsize{9}{9}\selectfont \textcolor{black}{(b)}}}
	\put(-90,33){{\fontsize{9}{9}\selectfont \textcolor{black}{(c)}}}
	\put(70,40){{\fontsize{9}{9}\selectfont \textcolor{black}{$\eta\leq1.75$}}}
	\put(70,30){{\fontsize{9}{9}\selectfont \textcolor{black}{$0-5\%$}}}
	\end{picture}
	\caption{(a) Correlator $r_{2,V}$ and its estimator $r_{2;\varepsilon_B}$ are obtained for symmetric collisions, Ne+Ne and O+O, at $0-5\%$ centrality. (b)  To check the estimations of flow decorrelations, the values of slop parameters $F_{2;V}$ (black) and $F_{2;\varepsilon_B}$ (red) have been shown.  Again, we note that ($L$) and ($R$) denote the collisions at LHC (6.37 TeV) and RHIC (200 GeV) energies.} 
	\label{fig11}
\end{figure}
\section{Initial state source}
As presented in Sec.\ref{sec3b}, we found that the decorrelations obtained from light nuclei collisions for pairs with a given $\Delta\eta$ are stronger at large rapidity than at mid-rapidity. This decorrelation of the flow vector is derived from the difference between $\mathcal{E}_{2,F}$ and $\mathcal{E}_{2,B}$, driven by the forward and backward going nucleons, respectively. This means that flow decorrelation can be estimated by its projection along $\mathcal{E}_{2,B}$ \cite{Zhang:2024bcb,Jia:2014ysa}:
\begin{equation}\label{q9}
v_{2,\varepsilon_B} (\eta)\equiv \frac{\la V_2(\eta)\mathcal{E}_{2,B}^*\ra}{\sqrt{\la\mathcal{E}_{2,B}\mathcal{E}_{2,B}^*\ra}}.
\end{equation}
Notice that $\mathcal{E}_{2,B}$ is obtained for $-3\leq\eta_{B}\leq-3.5$ at far-backward.
We compare this estimation with the 2-point correlation for a similar reference bin, $-3\leq\eta_{ref}\leq-3.5$:
\begin{equation}\label{q10}
v_{2,\eta_{ref}} (\eta)\equiv \frac{ v_2\{2\}^2(\eta,\eta_{ref})}{v_2\{2\}(\eta_{ref})}.
\end{equation}
Fig.\ref{fig11}a presents a comparison of the results of Eq.\ref{q9} with Eq.\ref{q10} for different system collisions. It indicates $v_{2,\eta_{ref}}>v_{2,\varepsilon_{B}}$ such that the ratio $v_{2,\eta_{ref}}/v_{2,\varepsilon_{B}}$ is equal to 1.3 for Ne+Ne, 1.1 for O+O(L) and 1.6 for O+O(R). This ratio is obtained 1.12 for $\eta<0$ and 1.3 for $\eta\geq0$ in d+Au collisions.
These results indicate that the correlator $v_{2,\eta_{ref}}$ approximately washes out the effect of asymmetry of the collisions, while $v_{2,\varepsilon_{B}}$ preserves it.   
Moreover, to present an estimation for the results depicted in Fig.\ref{fig9}, we study the correlator $r_{2;V}$ and its estimator, $r_{2;\varepsilon_B}=v_{2,\varepsilon_B}(\eta)/v_{2,\varepsilon_B}(-\eta)$. The results for $0-5\%$ central collisions are displayed in Fig.\ref{fig11}b. Although we have $r_{2}^{Ne+Ne}>r_{2}^{O+O(L)}>r_{2}^{O+O(R)}$, the difference between $r_{2}^{O+O(L)}$ and $r_{2}^{O+O(R)}$ is decreased concerning the projection of flow vector on $\mathcal{E}_B$. Nonetheless, this projection decreases the decorrelations as depicted in Fig.\ref{fig11}a. This results are consistent with the trends in Ref.\cite{Zhang:2024vkh}. It means that this tendency is also preserved in small systems. It also can be studied by employing slope parameter.  The results are shown in Fig.\ref{fig11}c.  It is observed that $F_{2;\varepsilon_B}<F_{2;V}$ and the relation  $F_{2}^{Ne+Ne}<F_{2}^{O+O(L)}<F_{2}^{O+O(R)}$ also exists for the estimation $F_{2;\varepsilon_B}$, although it cannot distinguish different collision energies in O+O.  However, this observable can highlight Ne+Ne collisions more significantly from O+O collisions. To check the estimations, the ratio $F_{2;\varepsilon_B}/F_{2;V}$ can be studied. It is obtained $\approx37\%$ for Ne+Ne, $\approx72\%$ for O+O(L) and $\approx65\%$ for O+O(R). Since the results of $r_2$ in Fig.\ref{fig11}a are not perfectly linear, we fit the right hand side of Eq.\ref{qFv} ($1-2\eta F_{2}$) to $r_2$. Fitting shows minor differences in the ratio of $F_{2;\varepsilon_B}/F_{2;V}$ as $\approx41\%$ for Ne+Ne, $\approx72\%$ for O+O(L) and $\approx61\%$ for O+O(R).  This means that the final state values can not be reached by the initial state estimation although the tendency is preserved.

\end{document}